\documentclass{article}

\usepackage{arxiv}

\usepackage[utf8]{inputenc} 
\usepackage[T1]{fontenc}    
\usepackage{hyperref}       
\usepackage{url}            
\usepackage{booktabs}       
\usepackage{amsfonts}       
\usepackage{nicefrac}       
\usepackage{microtype}      
\usepackage{lipsum}		
\usepackage{graphicx}
\usepackage{natbib}
\usepackage{doi}

\usepackage{algorithm}
\usepackage{algpseudocode}
\usepackage{comment}

\usepackage{hyperref}

\usepackage{calrsfs}
\usepackage{subfig}
\usepackage{amsmath}
\usepackage{amsfonts}
\usepackage{color}
\usepackage{rotating}

\newcommand{\TT}{\mathbb{T}}
\newcommand{\RR}{\mathsf{R}}
\newcommand{\ZZ}{\mathbf{Z}}
\newcommand{\Al}{\mathcal{A}}
\newcommand{\DD}{\mathcal{D}}
\newcommand{\VV}{\mathsf{V}}
\newcommand{\CC}{\mathsf{C}}
\newcommand{\Cset}{\mathbb{C}}
\newcommand{\Sc}{\mathsf{S}}

\newcommand{\Rset}{\mathcal{R}}
\newcommand{\Rule}{\mathsf{Rule}}
\newcommand{\View}{\mathsf{View}}
\newcommand{\Dest}{\mathsf{Dest}}
\newcommand{\cycle}{\vec{\mathcal{C}_f}}

\newcommand{\PRM}{\emph{Pairbot model}}
\newcommand{\PR}{\emph{pairbot}}
\newcommand{\SH}{\emph{short}}
\newcommand{\LN}{\emph{long}}
\newcommand{\bd}{\emph{buddy}}

\newcommand{\ASYNC}{\mathsf{ASYNC}}
\newcommand{\SSYNC}{\mathsf{SSYNC}}
\newcommand{\FSYNC}{\mathsf{FSYNC}}

\newcommand{\bud}{\mathsf{Buddy}}
\newcommand{\chk}{\mathsf{Chk}}
\newcommand{\nr}{\mathsf{Nr}}
\newcommand{\move}{\mathsf{Move}}

\newcommand{\ie}{{i.e., }} 
\newcommand{\etal}{{et al.}} 
\newcommand{\etc}{{etc.}} 
\newcommand{\eg} {{e.g., }} 

\newtheorem{theorem}{Theorem}
\newtheorem{lemma}{Lemma}
\newtheorem{corollary}{Corollary}

\newtheorem{proof}{Proof}
\newtheorem{definition}{Definition}

\title{Pairbot: A Novel Model for Autonomous Mobile Robot Systems Consisting of Paired Robots}

\author{ \href{https://orcid.org/0000-0002-5437-7626}{\includegraphics[scale=0.06]{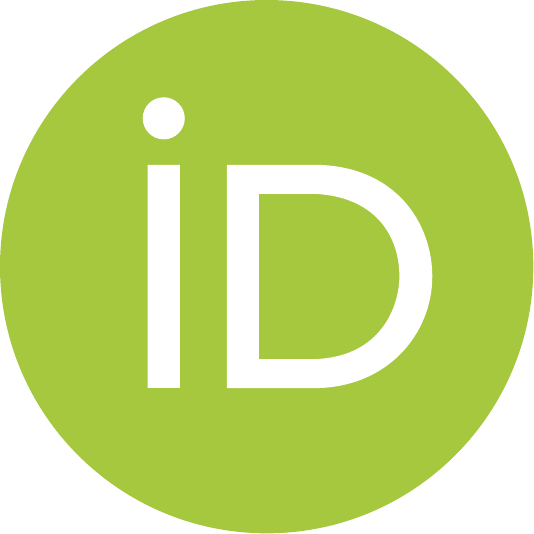}\hspace{1mm}Yonghwan Kim}\\
	Department of Computer Science\\
	Nagoya Institute of Technology\\
	Aichi, Japan \\
	\texttt{kim@nitech.ac.jp} \\
	\And
	\href{https://orcid.org/0000-0003-1683-2154}{\includegraphics[scale=0.06]{orcid.pdf}\hspace{1mm}Yoshiaki Katayama}\\
	Department of Computer Science\\
	Nagoya Institute of Technology\\
	Aichi, Japan \\
	\texttt{katayama@nitech.ac.jp} \\
	\And
	\href{https://orcid.org/0000-0002-5351-1459}{\includegraphics[scale=0.06]{orcid.pdf}\hspace{1mm}Koichi Wada}\\
	Department of Applied Informatics\\
	Hosei University\\
	Tokyo, Japan \\
	\texttt{wada@hosei.ac.jp} \\
}

\renewcommand{\shorttitle}{Pairbot: A Novel Model for Autonomous Mobile Robots}

\hypersetup{
pdftitle={Pairbot: A Novel Model for Autonomous Mobile Robot Systems Consisting of Paired Robots},
pdfsubject={q-bio.NC, q-bio.QM},
pdfauthor={Yonghwan Kim, Yoshiaki Katayama, Koichi Wada},
pdfkeywords={pairbot, programmable matter, autonomous mobile robots, LCM model},
}

\begin{document}
\maketitle

\begin{abstract}
\emph{Programmable matter (PM)} is a form of
matter capable of dynamically altering
its physical properties,
such as shape or density, through programmable means. 
From a robotics perspective, 
PM can be realized as a \textit{distributed system consisting of numerous small computational entities} 
working collaboratively to achieve specific objectives. 
Although autonomous mobile robot systems serve as an important example and have been researched for more than two decades, these robots often fail to perform even basic tasks, revealing a considerable gap in PM implementation.

In this paper, we introduce a novel computational paradigm, termed the Pairing Robot model (\PRM), 
which is built on an autonomous mobile robot system. In this model,
each robot forms a \emph{pair} with another, enabling them to recognize each other and
adapt their positions to achieve designated goals. 
This fundamental principle of \emph{pairing}
substantially enhances inter-robot connectivity compared to conventional \emph{LCM}-type model,
even under asynchronous scheduler conditions.
This shift has considerable implications for computational capabilities, specifically in problem solvability.

We explore two specific challenges--the \emph{perpetual marching} problem
and the \emph{7-pairbots-gathering} problem-- 
to demonstrate the computational power of \PRM.  
This model provides new avenues and insights to address inherent issues in autonomous mobile robots.

\end{abstract}


\section{Introduction}

\subsection{Background}
The foundational concept of a distributed system composed of multiple robots was initially introduced
in a landmark paper \cite{SY99}.
Within this framework, each robot autonomously observes the positions of the other robots and
moves to a new position based on a prescribed algorithm.
These robots are anonymous (\ie indistinguishable in 
 appearance) 
and uniform (\ie executing identical algorithms).
In the work cited~\cite{SY99}, this conceptual model for mobile robots
is designated as the \emph{LCM (Look-Compute-Move) model}. For simplicity,
we will refer to robots operating under the  conventional \emph{LCM} model
without specific assumptions such as  \emph{light} \cite{light1}) as \emph{LCM-robots}.
The paper provides a comprehensive analysis of the capabilities and constraints of this distributed system, exploring issues such as pattern formation and agreement problems.
Since the introduction of the \emph{LCM} model, 
much related work has been studied to clarify its computational power and limitations~\cite{DBLP:series/lncs/11340,mobilerobot} for more than 20 years.

Numerous studies have focused primarily on the relationship
between the computational capabilities of each robot
and the solvability of the given problem,
such as problems of gathering \cite{CDN,CFPS},
pattern formation \cite{pf2,pf1,FYOKY,YUKY}, and flocking \cite{GP,SIW}.

Thus, defining the necessary (possibly minimum) capabilities to solve the given problem 
is an essential issue.
Many capabilities of a robot (\eg geometric agreement, scheduler, visibility, and transparency) 
should be considered to solve a problem,
and it has been shown that the solvability of each problem depends deeply on these capabilities.
Clarifying the capabilities required for the given problem has many advantages, 
including cost reduction, scalability, and fault tolerance.
Therefore, various distributed coordination problems for autonomous mobile robots are still 
widely studied in many fields, such as robotics, engineering, and medical science.

\subsection{Related Work}

Building on the seminal work by Suzuki and Yamashita \cite{SY99}, 
which introduced the autonomous mobile robot model 
and explored its computational capabilities, 
extensive research has since been investigated concerning its computational power 
and limitations of these robots in various distributed coordination tasks,
such as rendezvous \cite{rend}, 
gathering \cite{CDN,CFPS}, 
pattern formation \cite{pf2,pf1,FYOKY}, 
dispersion \cite{disper0,disper1,disper2,scattering}, 
and flocking \cite{GP,SIW}.

In the early years of the study, 
robots are modeled as points (\ie without volume). 
However, many recent studies consider robots that have volume, known as \textit{ fat robots} \cite{fatrobot}, 
or opaque robots (each robot may obstruct the view of the other robots):
these models present new paradigms for problems such as
the complete visibility problem \cite{CV,CV2}. 
Many studies aim to clarify the relationship between robots' capabilities and the solvability of various problems \cite{DBLP:series/lncs/11340,mobilerobot,landscape}.

\emph{Programmable matter (PM)}, which is defined as a matter
that can change its physical properties,
such as shape, color, and stiffness, in response to a computer program,
was first introduced in \cite{pm93,pm}.
Realizing programmable matter could bring several potential advantages,
including:
(1) {\bf flexibility}; 
PM makes matters highly flexible and adaptable, which allows PM to be used in a variety of applications such as wearable devices and soft robotics \cite{sr1,sr2}, 
(2) {\bf customization}; 
PM could allow customization of objects and devices to meet specific needs and requirements, 
\eg tailored clothing or customized prosthetics,
and
(3) {\bf interactivity};
PM could bring new levels of interactivity and responsiveness to physical objects,
\eg objects that respond to touch, sound, or light in creative ways.
These potential benefits of PM are significant and could lead to many new and innovative applications. 
However, PM is a complex and challenging task that requires solving several technical and scientific problems,
and thus it is still in the research stage in various fields such as robotics, computer science, and material science.
PM can be implemented through the use of small robots, such as nanorobots \cite{nanorobot1,nanorobot2}, 
that can work together to form larger structures. 
Many studies on autonomous mobile robot systems have the potential to be useful in realizing PM.
However, autonomous mobile robots require various additional capabilities in many cases, and some of them are not necessarily realistic,
as they have many difficulties, even in simple problems.

As a new computational model for the realization of programming matter,
a self-organizing particle system, called \emph{Amoebot model}, was first introduced in \cite{amoebot}.
The \emph{Amoebot} model consists of a large number of computational \emph{particles} 
placed on a triangular grid plane,
which locally interact with each other to solve problems.
Each particle repeatedly changes its state to \emph{contraction} or \emph{expansion}, 
which means the state by which a particle occupies one node and two adjacent nodes.
The \emph{Amoebot} model allows for coordinated movement between two connected particles,
called \emph{handover}. 
A handover is an interaction in which a particle can contract out of a certain node
at the same time that another particle expands into that node.
The \emph{Amoebot} model can solve various distributed coordination problems, 
such as the universal coating problem \cite{sops1}, leader election \cite{LEamoe1,LEamoe2},
and convex hull formation \cite{sops3}.
Furthermore, recent work has introduced the \emph{canonical amoebot model},
which considers concurrent control \cite{camoebot}.

\emph{SILBOT} model \cite{silbot} is another computational model for PM
that introduces a new modeling approach by relaxing some of the assumptions made in the previous model.
In the \emph{SILBOT} model, each particle cannot communicate with the others (\ie silent)
and operates asynchronously (\ie in full synchronicity).
However, this model requires a specific symmetry-breaking capability;
if two or more particles attempt to expand toward the same cell, 
only one of them succeeds.
The literature \cite{silbot} shows that a leader election can be achieved
from a connected configuration with certain restrictions (or with some additional assumptions).

\emph{MOBLOT} \cite{moblot} is a computational model
designed from a different point of view,
which extends the oblivious mobile robot model to address a wider spectrum of cases. 
The main difference of the \emph{MOBLOT} model is heterogeneous system; 
all (oblivious) robots are partitioned into 2 types, colored black or white.
This feature enables simple (oblivious) robots to perform more complex tasks.

The literature \cite{camoebot,silbot,moblot} shows that these new computational models for PM
have great computational power in some problems;
however, they require some different assumptions
which are not considered (or cannot be used) in
the conventional autonomous mobile robot model,
such as explicit communication or handover.
Here, we expect that
if we can design a new computational model as close as possible to the
conventional \emph{LCM} model \cite{SY99}, 
we can use many results among many existing results investigated for a long time.

\subsection{Comparison with the Other Computational Models}
Here we provide a comparison between our proposed model \PRM~and three other computational models:
\emph{Amoebot}, \emph{SILBOT}, and \emph{MOBLOT}.
It is important to note that these models solve different problems
based on different assumptions; thus, it is difficult to provide a direct comparison of their computational power.

The \emph{Amoebot} model is one of the most popular PM models. 
Various \emph{Amoebot} models presented in many studies have recently been generalized in \cite{camoebot},
and it is currently one of the most promising models for PM.
It relies on explicit communication (message passing) between particles,
allowing them to move in coordinated ways, form complex structures, and perform various tasks.
The most notable feature of the \emph{Amoebot} model is the solvability of the leader election;
the communication between particles is a very useful capability to solve the leader election problem.
In many cases, the \emph{Amoebot} model uses the leader elected among the particles to solve the pattern formation problem \cite{SFamoe1,SFamoe2}.
The ability to elect the leader under the various assumptions is one of the strongest computational power of the \emph{Amoebot} model.

The \emph{SILBOT} model \cite{silbot} (as in \emph{Silent robot}) relaxes several assumptions, 
especially communication capacity.
G. D'Angelo \etal~show that a leader election,
at most three leaders could be elected due to the symmetry of the initial configuration,
can be achieved by this model without any communication
from any simple (\ie without holes) connected configuration.
As one of the features of \emph{SILBOT}, it assumes 2-hop visibility;
each particle can detect nodes within 2 hops.
To our knowledge, only the leader election problem is investigated under the \emph{SILBOT} model \cite{silbot},
however, this study shows that the leader election is solvable without any explicit communication under some specific model.
Although it is difficult to compare the computational power
between \emph{Amoebot} and \emph{SILBOT},
it is natural to say that \emph{Amoebot} is stronger than \emph{SILBOT} if there is no additional assumption.

The \emph{MOBLOT} model is a computational model based on the \emph{LCM} model;
thus, robots are silent (without explicit communication) and do not have special symmetry-breaking capabilities like \emph{SILBOT}.
Therefore, \emph{MOBLOT} is the model closest to our proposed model \PRM:
These two computational models are designed with the goal of significantly changing their computational power
by adding a few assumptions to conventional mobile robots.
\emph{MOBLOT} is a model inspired by molecules
in which oblivious autonomous mobile robots with some different roles
are combined to act as a new large computational entity;
such as atoms, molecules are formed.
The robots in the same molecules can operate together, \eg they have geometric agreement,
and this enables the robots to break their symmetry because the robots forming molecules become nonequivalent (\ie there is no symmetry).
The literature \cite{moblot} gives one special case study of pattern formation, named  matter formation, 
to form molecules by robots.
However, this study provides us with the fact that
this model can enlarge the class of achievable patterns of oblivious mobile robots that break the symmetry of the system.
Since \emph{MOBLOT} is based on conventional mobile robots, 
it is a relatively weak computational model compared to the two models above,but assumes relatively strong capabilities (even though they are not unrealistic) compared to our proposed model;
\PRM~assumes only the unique pair of each robot, which can be easily implemented (or simulated) with some previous assumptions, such as lights.

\emph{The Uniqueness of \PRM:} What separates \PRM\ apart is its minimalistic approach to computational capabilities, namely the unique pairing of each robot. The real strength of \PRM\ lies in its \emph{connectivity}, even under asynchronous schedulers, as shown in our solutions to the problems of \emph{perpetual marching} and \emph{7-pairbots-gathering}. The former illustrates that a problem unsolvable by LCM-robots under a semi-synchronous scheduler is, in fact, solvable by \PR s under an asynchronous scheduler. The latter shows how \PRM\ outperforms traditional robots in terms of visibility range. This clearly highlights the superior efficiency of \PRM in solving problems with fewer assumptions, offering advantages in terms of implementability and real-world applicability.

In summary, while \emph{Amoebot} offers the most computational power, followed by \emph{SILBOT} and \emph{MOBLOT}(these two models are hard to compare), \PRM\ carves out its own unique space by efficiently solving problems with fewer computational assumptions. This makes it an ideal choice for scenarios where computational simplicity and efficiency are critical.



\subsection{Contribution}
In this paper, we introduce a novel method for implementing programmable matter, 
designed to simplify the exploration and analysis of its capabilities using established knowledge.
Our model, known as \PRM, features two robots that operate collaboratively as a pair on a triangular grid.
The \PRM\ introduces a unique feature called an \emph{exclusive move},
which is absent from the traditional \emph{LCM} model.
Despite this addition, \PRM largely retains the functionalities found in the \emph{LCM} model.
This newly added capability significantly enhances  computational power, specifically problem 
 solvability,  
and offers valuable insights into the realization of  programmable matter based on the \emph{LCM} model. 

In \PRM, each robot is uniquely paired with another robot, referred to as \bd. 
We denote these paired robots as \PR, 
and a \PR\ system comprises  two or more \PR s.
Every robot in a \PR\ can identify its \bd. 
The paired robots repeatedly alter their geometric relationships, \SH\ and \LN, to achieve their objectives.
In the \SH~state, both robots share the same spatial point, while in the \LN state, they occupy adjacent points.

To elucidate the functioning of \PR, 
we present two challenges:
the \emph{perpetual marching} and the \emph{7-pairbots-gathering} problems. 
The former problem is not solvable by LCM-robots under a semi-synchronous scheduler,
and latter is not solvable by LCM-robots with visibility range 1.
We propose two deterministic algorithms as solutions for these challenges, 
which serve to demonstrate how \PR s operate and 
both the computational power and the limitations of the \PRM.
In particular, 
the first algorithm solves the perpetual marching problem
under an asynchronous scheduler,
and the second solves the 7-pairbots-gathering problem
by \PR s with visibility range 1.



\subsection{Paper Organization}
The rest of this paper is organized as follows:
Section \ref{sec:model} introduces the proposed system model, called \PRM; 
Section \ref{sec:marching} gives the \emph{perpetual marching} problem  
and proposes an algorithm to solve this problem; 
Section \ref{sec:7gather} discusses another problem, called the \emph{7-pairbots-gatheringproblem }
and an algorithm to solve this problem;
and finally Section 5 concludes the paper.


\section{Proposed Model: \PRM}
\label{sec:model}

\subsection{Triangular Grid Plane}

\begin{figure}[tbh]
  \centering
		\includegraphics[scale=1]{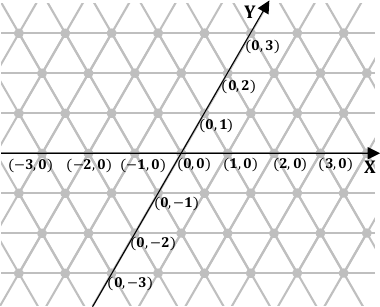}
  \caption{Triangular grid plane $\TT$}
  \label{tgrid}
\end{figure}

\begin{figure*}[bth]
  \centering
		\includegraphics[scale=0.8]{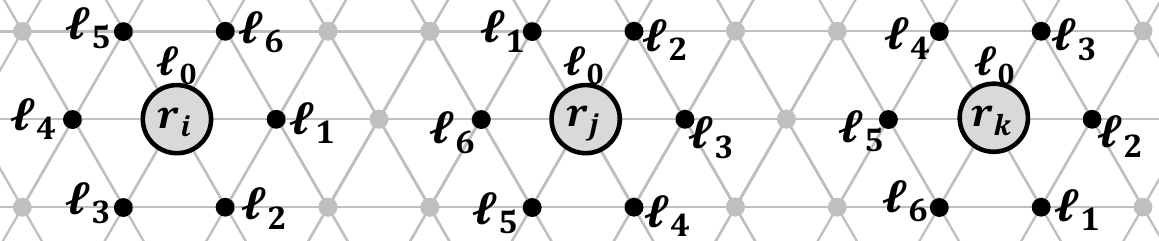}
  \caption{Example of the local labels of each robot without any geometric agreement}
  \label{fig:label}
\end{figure*}

We consider a set of $n$ autonomous mobile robots denoted by $\RR = \{r_1, r_2, \ldots, r_n\}$
on a two-dimensional triangular grid plane $\TT = \ZZ^2$ (Figure \ref{tgrid}).
The distance between two points $u$ and $v$ in $\TT$ is defined by the following equation:

\begin{eqnarray}
\mathit{dist}(u,v)=\left\{ \begin{array}{l}
    |u.x - v.x| + |u.y - v.y| \hspace{40pt} \mathit{\bf if}~ (u_x - v_x)(u_y - v_y) \geq 0 \\
    \mathit{max}(|u.x - v.x|,|u.y - v.y|) \hspace{20pt} \mathit{\bf otherwise} \\
\end{array} \right.
\end{eqnarray}

If the distance between two points is one, the two points are \emph{adjacent}.
The triangular grid plane $\TT$ can also be represented as an infinite regular graph $G_\TT = (V_\TT, E_\TT)$, 
where $V_\TT$ consists of all points on $\TT$ and $E_\TT$ is defined by all two adjacent points on $\TT$ 
(\ie for $u$, $v \in V_\TT$, $(u,v) \in E_\TT$ iff $\mathit{dist}(u,v) = 1$).

Note that we can use the function $\mathit{dist}()$ for robots such as $\mathit{dist}(r_i, r_j)$
which is equal to $\mathit{dist}(u,v)$ where $u$ (resp. $v$) is the point occupied by $r_i$ (resp. $r_j$).

\subsection{Geometric Agreement}
Every robot has its own local coordinate system that can be defined by \emph{directions} 
(\ie $X$ and $Y$ axes) and \emph{orientations} (\ie positive and negative sides) of each axis.
We can consider some levels of consistency among robots on their local compass:
\emph{total agreement}, when all robots agree on the directions and orientations of both axes;
\emph{partial agreement}, when all robots agree on the direction and orientation of only one axis or
on the \emph{chirality}, which means a sense of axis orientation
(\ie clockwise or counter-clockwise); 
or \emph{no agreement}, when no agreement exists among the local coordinate system of each robot.
In the problems considered in this paper, 
we assumed \emph{no agreement} and \emph{total agreement} respectively.
Note that the latter means that 
all robots in $\RR$ agree on the directions
and orientations of both axes, but no robot agrees on the position of the origin.
In other words, no robot knows its global coordinate, 
but all agree on the sense of direction (\eg north, south, east, and west).

\subsection{Pairbot}
In \PRM, every robot has its unique partner, called \bd: 
robot $r_i$ is the \bd\ of robot $r_j$ if and only if robot $r_j$ 
is the \bd\ of robot $r_i$.
In this case, we call the two robots $r_i$ and $r_j$ a \emph{pairbot}.
The \bd\ of each robot is initially determined and never changed.
Obviously, the number of robots $n$ is an even number. 

\subsubsection{Operations}
Each robot $r_i$ cyclically performs the following three operations: 
\emph{Look}, \emph{Compute}, and \emph{Move}
based on a well-known computational model (\emph{LCM} model \cite{SY99}).

\begin{itemize}
    \item \emph{Look}: Each robot takes a snapshot consisting of robots within the visibility range with respect to its local coordinate system.
    \item \emph{Compute}: Each robot performs a local computation based on the snapshot taken by the \emph{Look} phase according to Algorithm $\Al$. As a result of the \emph{Compute} phase, each robot determines its destination point to move.
    \item \emph{Move}: Based on the result of the \emph{Compute} phase, each robot actually moves to the adjacent destination point from the current point in the \emph{Move} phase.
    A null movement (staying) is allowed (\ie a robot does not move).
\end{itemize}

\subsubsection{Local Label}
Each robot locally maintains the labels for each incident edge 
(from $\ell_1$ to $\ell_6$) to distinguish its adjacent points.
Each robot selects one adjacent point 
and labels the point as $\ell_1$,
and labels the other points from $\ell_2$ to $\ell_6$ in clockwise order (based on its local chirality).
The label $\ell_0$ represents its current point.
Note that the direction located the point labeled by $\ell_1$ may be different 
by the assumption of geometric agreements among robots. 
Moreover, the clockwise order may also vary by the assumption of \emph{chirality}.

Figure \ref{fig:label} represents an example of the local labels of three robots, $r_i$, $r_j$, and $r_k$, 
where there is \emph{no agreement} among the robots. 
In this figure, robot $r_i$ labels the point on the right side as $\ell_1$; 
however, robot $r_j$ or $r_k$ labels the point at a different direction with $r_i$'s as $\ell_1$
because robots do not agree on the directions and orientation of any axis.
Moreover, the edges of robot $r_k$ are labeled in counterclockwise order, 
because neither do the robots agree on chirality.

In the problems discussed later in this paper, 
we assume either \emph{no agreement} or \emph{total agreement} depending on the problems.

\subsubsection{Capabilities}
All robots are oblivious (\ie do not have memory), 
which means that they do not know any of their past executions.

Each robot can recognize other robots within \emph{visibility range} $\VV$.
This means that each robot can sense only up to $\VV$ hops far from itself.
In other words, the robot $r_u$ at the point $u$ and the robot $r_v$ on point $v$, 
where $r_u, r_v \in \RR$ and $u, v \in V_\TT$,
can observe each other only when $\mathit{dist}(u,v) \leq \VV$.
We assume here that the \emph{visibility range} of each robot is one 
(\ie each robot can observe only the robots at the adjacent points).

We also assume that each robot has the capability of \emph{weak multiplicity detection}, that is,
each robot can distinguish from the following three cases at points within its visibility: 
no robot exists; only one robot exists; and more than one robot exists.
Note that if each robot can count the exact number of robots that occupy the point within its visibility range,
this is \emph{strong multiplicity detection}.

All robots in $\RR$ are \emph{anonymous}: 
each robot has no identifier and no robot is distinguishable by its appearance.
However, each robot maintains the position of its \bd\ using its local label, 
implying that each has a local memory to maintain the position (local label) where its \bd\ is located.

We further assume here that when a \PR~occupies the same point, 
only one robot can move to its adjacent point when it moves.
We call this \emph{an exclusive move}.
Note that no robot knows if any other two robots are \PR\ or not.

\begin{figure}[tbhp]
  \centering
		\subfloat[\SH~state]{\includegraphics[scale=1.2]{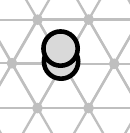}}\hspace{60pt}
		\subfloat[\LN~state]{\includegraphics[scale=1.2]{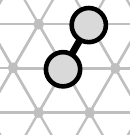}}
  \caption{Two states (positional relation) of two robots in the same pair}
  \label{fig:shortlong}
\end{figure}

The two robots in one \PR~occupy the same point 
or two adjacent points on $\TT$ (Figure \ref{fig:shortlong}).
We call the former one a \SH~state and the latter a \LN~state.
When a \PR~is in a \SH~state, only one robot in the \PR~can (exclusively) move 
the adjacent point from its current point, and the \PR~becomes a \LN~state. 
When a \PR~is in a \LN~state, either of the two robots can move to the point occupied by its \bd, 
and the state is changed back to a \SH~state.
In the \PRM, every \PR~repeatedly changes its state to \SH\ and \LN\ to achieve the goal.
A \PR\ never knows which robot moved while its state changes 
because all robots are oblivious.

\begin{figure}[tb]
  \centering
		\includegraphics[scale=0.8]{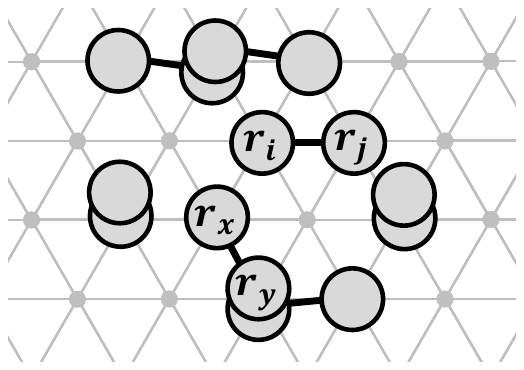}
  \caption{An example of \PR s}
  \label{fig:pairex}
\end{figure}

Figure \ref{fig:pairex} illustrates an example of \PR s.
\emph{Pairbot} $r_i$ and $r_j$ is in a \LN~state, 
and some other robots are placed at the other points in $\TT$.
In this case, robot $r_i$ recognizes that its \bd~is robot $r_j$ on its right side.
However, $r_i$ cannot know the pair relations of the other robots 
(\eg robot $r_i$ never knows the \bd~of robot $r_x$ on its lower left side).
The robot $r_y$ can recognize that another robot is at the same point 
due to \emph{weak multiplicity detection}. 
Furthermore, $r_y$ can know that the \bd~$r_x$ is on the point of its upper left side 
and is in a \LN~state.

\subsection{Scheduler}
We consider a \emph{scheduler} that decides which robot to \emph{activate} 
and the \emph{timing} of each operation.
A scheduler has three representative assumptions (models):
\emph{fully-synchronous} ($\FSYNC$); 
\emph{semi-synchronous} ($\SSYNC$); 
and \emph{asynchronous} ($\ASYNC$).

In an $\FSYNC$ scheduler, 
all robots are activated at the same time, 
and the three operations of \emph{Look}, \emph{Compute}, and \emph{Move}
are executed based on exactly the same cycle time.
In an $\SSYNC$ scheduler, all robots perform their operations at the same time, 
but some may not be activated. 
The robots, which are not activated by a scheduler, wait until all activated robots 
terminate their operations.
Lastly, in an $\ASYNC$ scheduler, no assumption on the cycle time of each robot is provided,
implying that all robots execute their operations at unpredictable time instants and durations.

We assume herein an $\ASYNC$ scheduler such that the \emph{Move} phase operates atomically:
each robot requires an unpredictable finite time to operate in the \emph{Look} or
\emph{Compute} phase, but it can atomically move in a property called the \emph{move-atomic property}.
Therefore, each robot is never observed while moving.
We assume that the two robots, which are a \PR\ are activated at the same time by the scheduler.

\subsection{Configuration}
A \emph{configuration} $\CC_t$ consists of the positions of all robots in $\RR$ at time $t$:
$\CC_t = \{ (r_1.x(t),$ $r_1.y(t)), (r_2.x(t), r_2.y(t)), \ldots, (r_n.x(t), r_n.y(t)) \}$, 
where $(r_i.x(t), r_i.y(t))$ is the global coordinate of robot $r_i$ at time $t$ on $\TT$.
Note that no robot knows its global coordinate on $\TT$.
Let $\RR'$ be a non-empty subset of $\RR$ and let $\Al$ be an algorithm.
We denote $\CC_t \mapsto_{(\RR', \Al)} \CC_{t+1}$ if a configuration $\CC_{t+1}$ is obtained 
when each robot in $\RR'$ simultaneously performs its \emph{Move} operation of $\Al$ in configuration $\CC_t$.
Hence, a scheduler can be presented as an infinite sequence $\RR_1, \RR_2, \cdots$ of nonempty subsets of $\RR$.
An execution $\Xi_\Al(\Sc,\CC_0)$ of Algorithm $\Al$ along the schedule $\Sc=\RR_1,\RR_2,\cdots$
starting from configuration $\CC_0$, where $\CC_0$ is an initial configuration, 
can be defined as the infinite sequence of configurations $\CC_0, \CC_1, \cdots$ 
such that $\CC_i \mapsto_{(\RR_{i+1}, \Al)} \CC_{i+1}$ for all $i \geq 0$.

\section{The Perpetual Marching Problem}
\label{sec:marching}

\subsection{Problem Definition}
In this section,
we consider the \emph{perpetual marching problem},
which is an infinite linear movement of \PR s without any geometrical agreement 
(\ie they do not have any common sense of direction).

We now define the marching problem as follows:



\begin{definition}
\emph{\bf{Perpetual Marching Problem.}} 
Algorithm $\Al$ solves the perpetual marching problem 
if there is a well-initiated configuration $\CC_{init}$, 
a certain direction $\DD$, 
a constant $k$,
and 
an infinite execution $\Xi_\Al(\Sc, \CC_{init})$ 
such that 
after each (constant) $t$ steps,
all robots move to a point that is a distance of $k$ toward the direction $\DD$.
\end{definition}

\begin{figure}[tb]
  \begin{center}
   \includegraphics[scale=0.7]{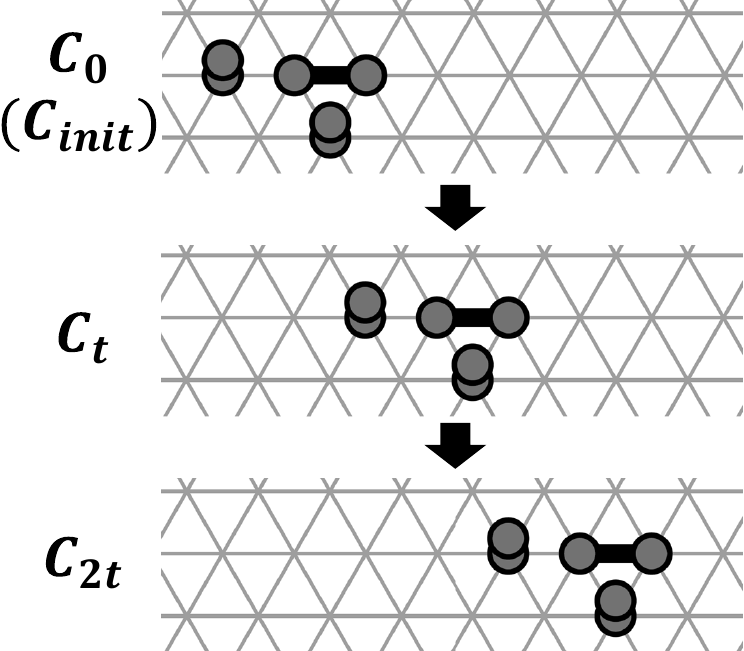}
  \end{center}
  \caption{An example of perpetual marching}
  \label{fig:pm-ex}
\end{figure}

Intuitively, when locating the robots in a predetermined configuration (a well-initiated configuration $\CC_{init}$),
an algorithm makes all the robots to move infinitely (an infinite execution) 
at a constant speed (a constant $k$) in a specific direction (a certain direction $\DD$).
Figure \ref{fig:pm-ex} illustrates an example of perpetual marching.

\begin{figure}[tb]
  \centering
		\subfloat[]{\includegraphics[scale=0.55]{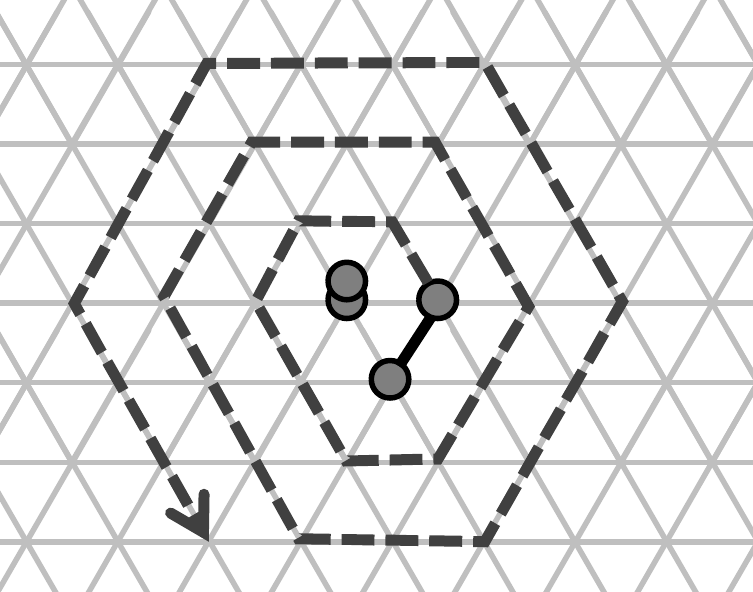}} \hspace{10pt}
		\subfloat[]{\includegraphics[scale=0.55]{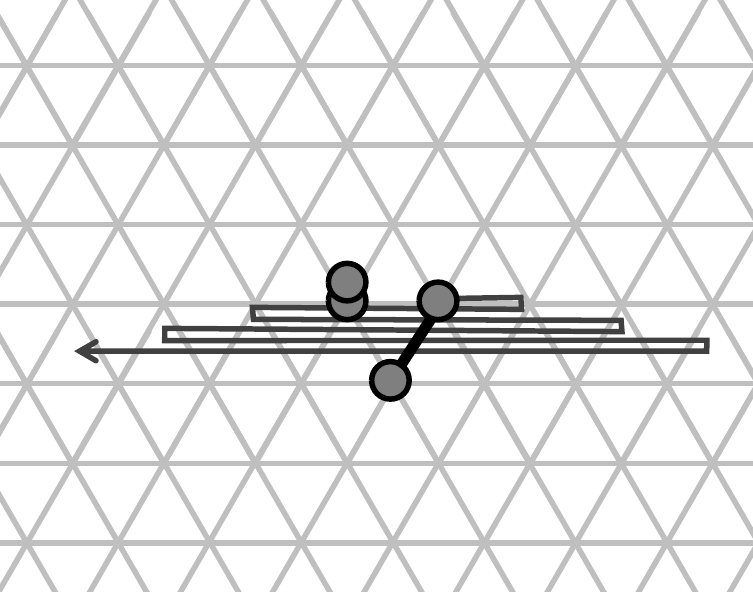}}
  \caption{Examples of non-feasible perpetual marching}
  \label{fig:nonmarching}
\end{figure}

\begin{figure}[tbhp]
  \centering
		\subfloat[]{\includegraphics[scale=0.55]{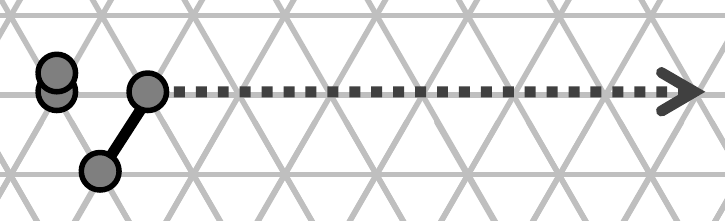}}
  
		\subfloat[]{\includegraphics[scale=0.55]{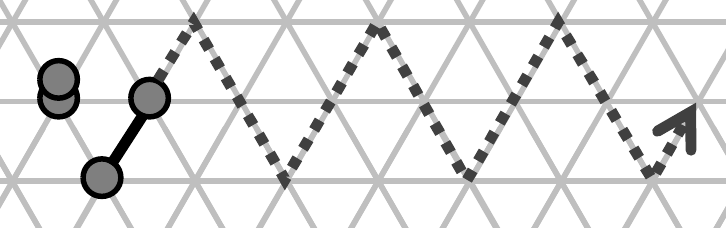}}
  \caption{Examples of feasible perpetual marching}
  \label{fig:pmex}
\end{figure}

It is worthwhile to noting that 
the perpetual marching requires a fixed distance movement within a constant speed. 
In other words, a movement in a specific direction within a finite time 
is not feasible perpetual marching. 
For instance, a movement such as a perpetual exploration while expanding the range 
(refer to Figure \ref{fig:nonmarching}(a))
or a repetitive movement back and forth 
(refer to Figure \ref{fig:nonmarching}(a))
is not feasible in the perpetual marching problem,
because from the perspective of moving in a fixed direction, 
their speed progressively decreases. 
On the other hand, whether the movement is in a straight line 
(refer to Figure \ref{fig:pmex}(a))
or follows a zigzag path, 
(refer to Figure \ref{fig:pmex}(b))
it becomes a feasible perpetual marching as long as it maintains a constant speed.

\subsection{Perpetual Marching Algorithm}
\label{sec:permarching}

\begin{figure}[tbhp]
  \begin{center}
   \includegraphics[scale=7]{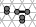}
  \end{center}
  \caption{An well-initiated Configuration $\CC_{init}$}
  \label{c_init}
\end{figure}

\begin{figure}[tbhp]
  \centering
		\includegraphics[scale=1]{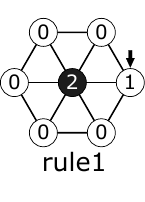}
		\includegraphics[scale=1]{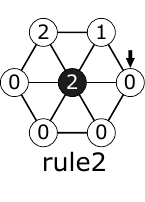}
		\includegraphics[scale=1]{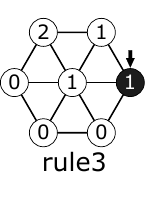}
		\includegraphics[scale=1]{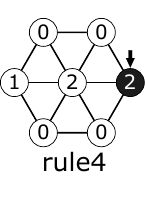}	

        \includegraphics[scale=1]{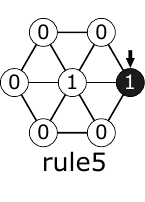}
		\includegraphics[scale=1]{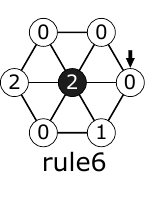}
		\includegraphics[scale=1]{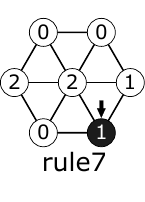}
  \caption{The proposed algorithm for perpetual marching}
  \label{pmalg}
\end{figure}

First we introduce an well-initiated configuration $\CC_{init}$ as Figure \ref{c_init}.
Figure \ref{pmalg} shows the proposed algorithm (using graphical representation) consisting of 7 rules 
when configuration $\CC_{init}$ is given.
The detailed pseudocode of the algorithm is provided in the next section.
In each rule, 
the circle in the center represents the grid point where a robot (who executes the algorithm) exists 
and 6 other circles present adjacent grid points.
Each number shows the number of robots 
and a black circle represents the point where the \bd~exists.
It is worth to noting that 
even if there are three or more robots at one point, 
the number (of robots) becomes 2 due to the weak multiplicity.
Moreover, 
these rules are applied as the same rules even mirrored and/or rotated
because robots do not agree on any axis nor chirality.
If the observation result of the robot satisfies any of these conditions, 
the robot moves to the point indicated by a black arrow.
Note that the direction to move is uniquely determined in each rule, 
thus every robot can move to the correct direction without a sense of direction.

During the execution of the algorithm,
eight different configurations appear sequentially.
After 8 steps,
the initial configuration appears again with a shift of distance one to the right direction.
These 8 steps are infinitely repeating and the perpetual marching is achieved.
As a result, the following corollary holds.

\begin{lemma}
\label{lem:marchingcorrect}
    The proposed algorithm solves the perpetual marching problem 
    under an $\ASYNC$ scheduler.
\end{lemma}

\begin{proof}
In configuration $\CC_{init}$,
there exists only one \PR~satisfying a rule (rule 1).
Therefore, even under an $\ASYNC$ scheduler, 
only one \PR~can move.
After one step, another configuration appears; however, in this configuration,
only one \PR~satisfies a rule (rule 2).
In all subsequent configurations, 
only one \PR~can move.
After 8 steps,
a configuration that is a translation of the configuration $\CC_{init}$ appears, 
which means that all robots move to the adjacent point at the same direction.
Figure \ref{exalg} shows the robot that satisfies a rule (a white robot),
and the rule number can be applied in each configuration.

\begin{figure}[tb]
  \centering
		\includegraphics[scale=0.9]{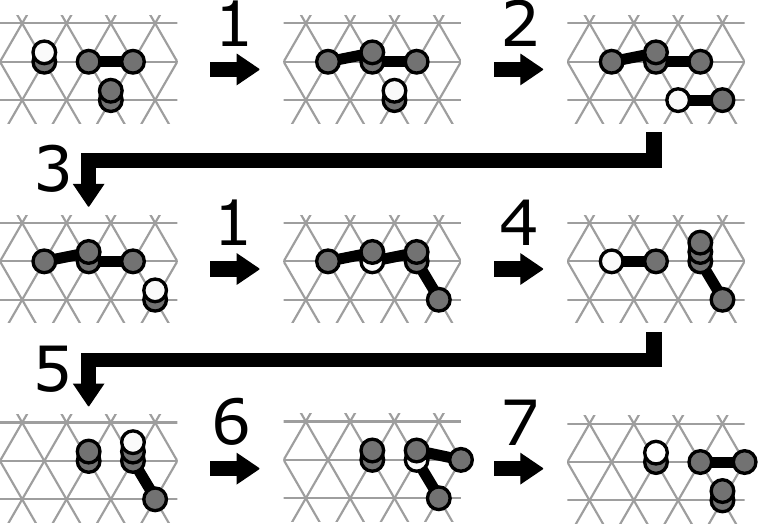}
  \caption{Execution of the proposed algorithm}
  \label{exalg}
\end{figure}

As a result, in every configuration, 
the proposed algorithm moves only one robot, and eight different configurations sequentially appear
while moving one hop in a certain direction.
This implies that the proposed algorithm solves the perpetual marching problem.
\end{proof}

By the proof of Lemma \ref{lem:marchingcorrect}, the following corollary holds.

\begin{corollary}
\label{cor:marchingstep}
    The proposed algorithm moves all the robots to a predetermined 
    direction (by an initial configuration) with a distance of 1 in 8 steps.
\end{corollary}

Here the term \emph{steps} means the total number of robots that moves.
By Lemma \ref{lem:marchingcorrect} and Corollary \ref{cor:marchingstep},
the following theorem holds.

\begin{theorem}
    When a well-initiated configuration is given,
    the proposed algorithm solves the perpetual marching problem
    by moving all the robots to the same direction with a distance of 1 
    in every 8 steps  
    under an $\ASYNC$ scheduler.
\end{theorem}

Note that 
the proposed algorithm is optimal in terms of the number of \PR s;
impossibility results for 1 \PR~or 2 \PR s can be easily proven 
because the total number of (feasible) configurations is small enough.

\subsection{Pseudocode of the Perpetual Marching Algorithm}
\label{sec:apa}
In this section, we present the pseudocode of the proposed algorithm 
to solve the perpetual marching 
by \PR s without a sense of direction.

Algorithm \ref{alg:pmarching} shows the proposed algorithm.
The proposed algorithm consists of 7 rules;
each rule is presented as \textsf{[Rule No.]: [Condition] $\to$ [Action]},
which means that a robot executes an \textsf{[Action]} 
when \textsf{[Condition]} in the same \textsf{[Rule No.]} is satisfied.
Since robots do not agree on any axis nor chirality,
all conditions in the algorithm are \textbf{rotatable} and \textbf{reflectable};
which means that 
the condition for each rule becomes \textsf{TRUE} if the observed result of the robot satisfies
the condition when rotated and/or mirrored in any direction.

\begin{algorithm}[thbp]
    \caption{Algorithm for perpetual marching}
    \label{alg:pmarching}
    {\bf variables and functions:}
        \begin{description}
        \item[$~\cdot~ \bud \in \{\ell_0, \ell_1, \ldots, \ell_6\}$~:] A label where its \bd~exists.
        \item[$~\cdot~ \chk(\ell_x, \nr)$~:] A function for checking the number of robots
        that returns $\mathsf{TRUE}$ when 
        the number of robots on the adjacent point at $\ell_x$ is the same as $\nr$. 
        Note that $\nr \in \{0,1,2\}$ due to weak multiplicity detection.
        The first parameter can be a set of labels $\mathbf{L} \subseteq \{\ell_x | 0 \leq x \leq 6 \}$; 
        the function returns $\mathsf{TRUE}$ only when $\chk(\forall \ell_x \in \mathbf{L}, \nr) = \mathsf{TRUE}$.
        \item[$~\cdot~ \move(\ell_x)$~:] Move to $l_x$.
\end{description}
\vspace{3pt}
{\bf algorithm:}
    \begin{algorithmic}
\State \textbf{/*} All conditions are \textbf{rotatable} and \textbf{reflectable}:
\State \hspace{10pt} the condition for each rule becomes \textsf{TRUE} if the observed result of the robot satisfies
\State \hspace{10pt} the condition when rotated and/or mirrored in any direction.\textbf{*/}
\State Rule1: $\bud=\ell_0 \land \chk(\{\ell_2,\ell_3,\ell_4,\ell_5,\ell_6\},0) \land \chk(\ell_1,1) \to \move(\ell_1)$
\State Rule2: $\bud=\ell_0 \land \chk(\{\ell_1,\ell_2,\ell_3,\ell_4\},0) \land \chk(\ell_5,2) \land \chk(\ell_6,1) \to \move(\ell_1)$
\State Rule3: $\bud=\ell_1 \land \chk(\{\ell_2,\ell_3,\ell_4\},0) \land \chk(\{\ell_0,\ell_1,\ell_6\},1) \land \chk(\ell_5,2) \to \move(\ell_1)$
\State Rule4: $\bud=\ell_0 \land \chk(\{\ell_2,\ell_3,\ell_5,\ell_6\},0) \land \chk(\ell_1,2) \land \chk(\ell_4,1) \to \move(\ell_1)$
\State Rule5: $\bud=\ell_1 \land \chk(\{\ell_2,\ell_3,\ell_4,\ell_5,\ell_6\},0) \land \chk(\{\ell_0,\ell_1\},1) \to \move(\ell_1)$
\State Rule6: $\bud=\ell_0 \land \chk(\{\ell_1,\ell_3,\ell_5,\ell_6\},0) \land \chk(\ell_2,1) \land \chk(\ell_4,2) \to \move(\ell_1)$
\State Rule7: $\bud=\ell_0 \land \chk(\{\ell_3,\ell_5,\ell_6\},0) \land \chk(\ell_2,1) \land \chk(\ell_4,2) \to \move(\ell_2)$
    \end{algorithmic}
\end{algorithm}

\subsection{Impossibility Result of Perpetual Marching by LCM-Robots}
\label{sec:apc}
Now we discuss the \emph{perpetual marching} problem also in \emph{LCM} model,
with no geometric agreement (\ie without a sense of directions).
If we assume an $\FSYNC$ scheduler, 
the problem can be easily solved by 3 autonomous mobile robots with visibility range 1
using the following algorithm.

\begin{itemize}
    \item Initial configuration $\CC_{init}$: Let $u$ and $v$ be the two adjacent points.
    The robot $r_1$ is located $u$ and the robots $r_2$ and $r_3$ are located $v$.
    \item Algorithm: If a single robot observes two accompanied robots, it moves to the point occupied by them. If an accompanied robot observes a single robot,
    it moves to the point opposite to it. 
\end{itemize}

Two pivotal questions naturally emerge:
\begin{enumerate}
\item \emph{"Is the perpetual marching problem solvable by LCM-robots under either an $\SSYNC$ or an $\ASYNC$ scheduler?"}
\item \emph{"What is the minimum number of LCM-robots required to solve the perpetual marching problem in the \emph{LCM} model under either an $\SSYNC$ or an $\ASYNC$ scheduler?"}
\end{enumerate}

To partially address these queries, we show the following theorem in this section. 

\begin{theorem}
\label{thm:impos}
Under an $\SSYNC$ scheduler, no deterministic algorithm exists that can solve the perpetual marching problem using unoriented (\emph{i.e.,} lacking geometric agreement) LCM-robots with a visibility range of 1, provided the number of robots is six or fewer.
\end{theorem}

Now we prove Theorem \ref{thm:impos} in the following subsections.

\subsubsection{Preliminaries}
Let $\RR = \{r_1, r_2, \ldots, r_6\}$ be a set of 6 autonomous mobile robots 
with visibility range 1.
The robots do not have any geometrical agreement (\ie they do not agree on any axis nor chirality).
Moreover, we assume that each robot has the capability of weak multiplicity detection.
Here we consider an $\SSYNC$ scheduler.

\begin{figure}[tbhp]
  \centering
		\subfloat[1-point configuration]{\includegraphics[scale=1.5]{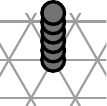}}\hspace{20pt}
		\subfloat[2-point configuration]{\includegraphics[scale=1.5]{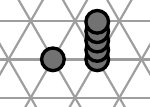}}\hspace{20pt}
        \subfloat[2-point configuration]{\includegraphics[scale=1.5]{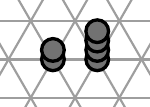}}\\
        
		\subfloat[2-point configuration]{\includegraphics[scale=1.5]{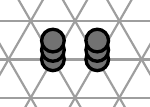}}\hspace{40pt}
        \subfloat[3-point configuration]{\includegraphics[scale=1.5]{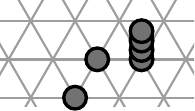}}\\
        
		\subfloat[3-point configuration]{\includegraphics[scale=1.5]{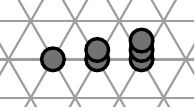}}\hspace{40pt}
        \subfloat[3-point configuration]{\includegraphics[scale=1.5]{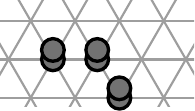}}\\
        
		\subfloat[4-point configuration]{\includegraphics[scale=1.5]{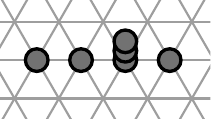}}\hspace{40pt}
        \subfloat[4-point configuration]{\includegraphics[scale=1.5]{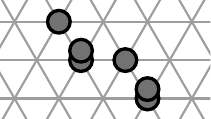}}\\
        
		\subfloat[5-point configuration]{\includegraphics[scale=1.5]{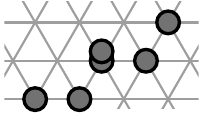}}\hspace{40pt}
        \subfloat[6-point configuration]{\includegraphics[scale=1.5]{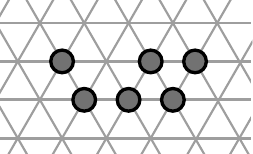}}\\
  \caption{Examples of $k$-point configurations}
  \label{fig:kconfex}
\end{figure}

A configuration $\CC_i$ consists of all positions of all robots in $\RR$,
and we call two configurations $\CC_i$ and $\CC_j$ is different 
when $\CC_i$ cannot be transformed into $\CC_j$ through translation, mirroring, and/or rotation.
We consider a set of all \emph{connected} configurations consisting of 6 autonomous mobile robots 
$\Cset = \{\CC_1, \CC_2, \CC_3, \ldots, \CC_n\}$
such that configurations $c_i$ and $c_j$ for any $i$ and $j$ ($i \neq j$) are different configurations.
A configuration $\CC_i$ is a $k$-point configuration
if exactly $k$ points are occupied in $\CC_i$.
Let $\Cset^k$ be a set of all $k$-point configuration in $\Cset$;
$\Cset^1 \cup \Cset^2 \cup \Cset^3 \cup \Cset^4 \cup \Cset^5 \cup \Cset^6 = \Cset$ and
$\Cset^1 \cap \Cset^2 \cap \Cset^3 \cap \Cset^4 \cap \Cset^5 \cap \Cset^6 = \emptyset$.
Figure \ref{fig:kconfex} shows the examples of $k$-point configurations by 6 robots.

Algorithm $\Al$ consists of a set of rules;
a rule $\Rule_i$ is defined as a tuple ($\View_i$,$\Dest_i$) 
which means that a robot moves to $\Dest_i$ if its observation result is the same as $\View_i$.
Note that the number of the possible observation results is finite (because the visibility range is assumed to 1),
the number of the possible actions is also finite (because each robot can move to a point among its 6 adjacent points).
Let $\Rset = \{\Rule_1, \Rule_2, \ldots, \Rule_m\}$ be a set of all possible rules.
Algorithm $\Al$ can be defined as a set of (non-conflicted) rules $\Rset_\Al \subset \Rset$.

Now we consider an algorithm-based directed graph $\overrightarrow{G_\Al}=(V,A_\Al)$ where $V = \Cset \cup \{\CC_{disCon}\}$.
Arc (\ie directed edge) $(\CC_i, \CC_j) \in E_\Al$ 
if $\CC_j$ is obtained when a robot performs its \emph{Move} operation according to a rule in algorithm $\Al$ in configuration $\CC_i$;
$\CC_i \mapsto_{(r_k, \Al)} \CC_j$ where robot $r_k$ satisfies a rule in $\Al$.
Moreover, arc (\ie directed edge) $(\CC_i, \CC_{disCon}) \in E_\Al$ 
if a non-connected configuration is obtained when a robot executes a rule in $\Al$ in configuration $\CC_i$

\subsubsection{Basic Strategy of the Proof}
First, we present the basic strategy of the proof. 
We assume an $\SSYNC$ scheduler, 
this means that in each round, the scheduler determines a set of robots and they executes the (same) algorithm.
If there exists an algorithm to solve the perpetual marching problem, it must be able to handle all patterns from all possible schedulers. 
Additionally, since the robots do not have a sense of direction, 
the destination point a robot moves to may not be uniquely determined
depending on the rule in the algorithm.
An algorithm for the perpetual marching problem must work correctly even 
even when the destination is arbitrarily selected. 
Therefore, in the proof of impossibility, 
we assume that both the scheduler and the destination are chosen adversarially, 
and show that there is no algorithm can solve the perpetual marching problem.

In the proof of impossibility, we assume the following adversarial scenarios:

\begin{enumerate}
    \item When there are two robots satisfying $\View_i$ and if they simultaneously execution of $\Rule_i$ makes the perpetual marching impossible, 
    the scheduler selects these two robots and makes them act in a way that makes the perpetual marching impossible.
    \item Except for \textbf{1}, the scheduler always selects only one point where there exists a robot satisfies any rule in the algorithm,
    and make all the robots at the point simultaneously execute the algorithm (\ie the scheduler never selects two different robots at two different points).
    \item In the case of \textbf{2}, when the destination is arbitrarily selected, 
    all robots choose the same direction. 
    In other words, if multiple robots are on the same point, they always move as a group and never scatter during execution.
    \item If a destination is not uniquely determined by a rule, 
    the scheduler selects destinations in a way that results in unfavorable configuration transitions
    (\eg if a specific movement of some robots causes a disconnected configuration, the scheduler selects that destination).
\end{enumerate}

We generate digraph $\overrightarrow{G_\Al}$ as a configuration transition diagram under the above scenario to prove the impossibility.
If there is algorithm $\Al$ to solve the perpetual marching problem, 
digraph $\overrightarrow{G_\Al}$ includes a directed cycle (denote $\cycle = (\CC_i, \CC_j, \CC_k, \ldots, \CC_i)$)
presenting an infinite execution (\ie infinite transition of configurations) of algorithm $\Al$.

Remind that every robot's visibility range is 1
and the robots do not agree on any axis nor chirality,
thus a connected configuration cannot be obtained from any disconnected configuration 
due to an adversarial scheduler.
Therefore, the following corollary holds.

\begin{corollary}
\label{cor:discon}
    If there exists directed cycle $\cycle$ in digraph $\overrightarrow{G_\Al}$,
    $\cycle$ consists of only connected configurations.
\end{corollary}

We call 
an infinite execution of algorithm $\Al$ from an well-initiated configuration 
(\ie any configuration appeared in $\cycle$)
\emph{the execution of} $\cycle$.

Moreover, 
we call directed cycle $\cycle$ in digraph $\overrightarrow{G_\Al}$ is \emph{illegible},
if $\cycle$ satisfies any of the following conditions.

\begin{enumerate}
    \item \textbf{(No movement)} Let $\CC_i$ be a well-initiated configuration in $\cycle$. 
    After the constant number of executions, when $\CC_i$ appears again (denote as $\CC_i^2$), 
    all the robots are at the exactly same points in $\CC_i$ and $\CC_i^2$. 
    \item \textbf{(Faulty concurrency)} In configuration $\CC_i$ in $\cycle$, 
    there are two different robots ($r_x$ and $r_y$) and two different rules ($\Rule_x$ and $\Rule_y$ in $\Al$)  
    such that robot $r_x$ (resp. $r_y$) can execute $\Rule_x$ (resp. $\Rule_y$). 
    When the two robots simultaneously execute $\Al$, 
    configuration $\CC_j$ which is not included in $\cycle$ is obtained.
    \item \textbf{(Adversarial transition)}
    Let $\CC_i$ be a configuration in $\cycle$,
    and $\CC_j$ be a configuration NOT in $\cycle$.
    There exists arc $(\CC_i, \CC_j)$ in $A_\Al$.
\end{enumerate}

\subsubsection{Proof of Impossibility Result}
Let $\Al$ be an algorithm to solve the perpetual marching problem 
by autonomous mobile robots with visibility range 1
under an $\SSYNC$ scheduler.

We call a rule such that
when a robot executes the rule,
all the robots located at seven points within the visibility range of the robot 
become disconnected 
a \emph{locally disconnecting rule}.

\begin{figure}[tbhp]
  \centering
		\subfloat[]{\includegraphics[scale=1]{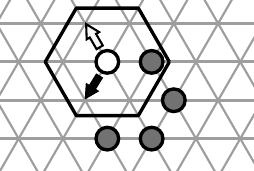}}\hspace{5pt}
		\subfloat[]{\includegraphics[scale=1]{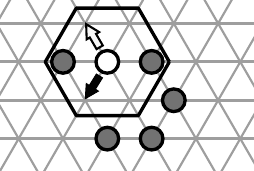}}\hspace{5pt}
        \subfloat[]{\includegraphics[scale=1]{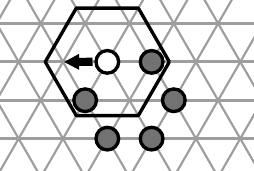}}
  \caption{Examples of locally disconnecting rules}
  \label{fig:disrules}
\end{figure}

Figure \ref{fig:disrules} shows some examples of locally disconnecting rules.
In each figure, a hexagon represents a local view of the observing robot (white circle) 
and each arrow shows a destination to move when the robot obtains such an observation result.
Depending on the configuration, 
even if some rule (\eg Figure \ref{fig:disrules}(c)) is executed, 
there are cases where the next configuration remains connected. 
However, in some configurations, 
it is possible for it to become disconnected. 
Since the robots executing the rules cannot perceive configurations 
outside their visibility range, 
adding locally disconnecting rules to the algorithm $\Al$ may result in execution 
even in cases where it becomes disconnected, leading to disconnected configurations.
From the above fact and Corollary \ref{cor:discon}, the following lemma holds.

\begin{lemma}
\label{lem:nolocaldis}
There is no locally disconnecting rule in algorithm $\Al$.
\end{lemma}

Now we have prepared the groundwork to proceed with the proof of impossibility. 
We show the impossibility by showing that none of the vertices in the graph are included in $\cycle$. Specifically, we start by proving that none of the points in all 1-point configurations
($\in \Cset^1$) are included, 
and then consider the points in all 2-point configurations ($\Cset^2$). 
In any 2-point configuration, 
we utilize the fact that the algorithm does not contain rules 
that transform the configuration into any 1-point configuration 
to reduce the number of vertices to consider. 
This is achieved by ensuring 
that when two robots at different points move to the same point once, 
the scheduler always activates these two robots simultaneously, 
providing the same view and ensuring that they always perform the same operations
(remind that we consider a deterministic algorithm).
This implies that if configuration $\CC_i$ in $\Cset^k$ is obtained 
from configuration $\CC_j$ ($i \neq j$) in $\Cset^{(k+1)}$ by an algorithm,
configuration $\CC_i$ cannot be obtained again by the algorithm 
due to an adversarial scheduler.
Hence the following corollary holds.

\begin{corollary}
\label{cor:samesetk}
If there exists 
directed cycle $\cycle$ (\ie algorithm to solve the perpetual marching exists),
all the configurations included in $\cycle$
are the same $k$-point configurations (\ie $\forall \CC_i , \CC_j \in \cycle: \{\CC_i , \CC_j\} \subseteq \Cset^k$).
\end{corollary}



\begin{figure}[t]
 \begin{minipage}{0.45\hsize}
  \centering
		\includegraphics[scale=2]{c1_1.pdf}
  \caption{Unique configuration $\CC_1^1$ in $\Cset^1$}
  \label{fig:cset1}
 \end{minipage}
 \hfill
 \begin{minipage}{0.45\hsize}
  \centering
		\includegraphics[scale=1.5]{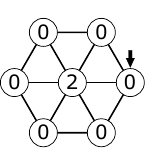}
  \caption{Only one feasible rule in $\CC_1^1$}
  \label{fig:rule1-1}
 \end{minipage}
\end{figure}

From the definition of $k$-point configuration, there is only one configuration included in $\Cset^1$.
Figure \ref{fig:cset1} presents the unique configuration of 1-point configuration ($\Cset^1$).

\begin{lemma}
\label{lem:noc1}
Let $\CC_1^1$ be the unique configuration in $\Cset^1$.
Cycle $\cycle$ never includes $\CC_1^1$.
\end{lemma}

\begin{proof}
Configuration $\CC_1^1$ is the same as the configuration depicted in Figure \ref{fig:kconfex}(a).
This implies that all the robots have the same observation result.
Figure \ref{fig:rule1-1} represents the observation result of every robot in $\CC_1^1$ 
(the notation of the rule is the same as the graphical representation of the algorithm
for perpetual marching;
refer to Section \ref{sec:permarching}).
The destination of the rule presented in Figure \ref{fig:rule1-1}
is the point on the right,
however, the destination point is irrelevant in this rule. 
In other words, this means that the rule remains the same 
regardless of where the destination point is
because the robots do not agree any axis.
We call this rule $\Rule_1$

Assume that configuration $\CC_1^1$ is included in $\cycle$.
If algorithm $\Al$ includes $\Rule_1$,
the scheduler activates all the 6 robots simultaneously, 
and makes them to move to the same point.
After that, the scheduler activates all the robots again,
and makes them to return back to the previously located point.
By repeating this, no robot cannot move to any other point than these two points.
As a result, the perpetual marching cannot be solved.
If algorithm $\Al$ does not include $\Rule_1$,
no robot cannot move.
Therefore, there is no algorithm to solve the perpetual marching problem 
from configuration $\CC_1^1$,
and this implies that configuration $\CC_1^1$ is not included in $\cycle$.
\end{proof}

\begin{lemma}
\label{lem:noc2}
Cycle $\cycle$ never includes any configuration in $\Cset^2$.
\end{lemma}

\begin{proof}
\begin{figure}[tbhp]
  \centering
		\subfloat[Configuration $\CC_1^2$]{\includegraphics[scale=1.5]{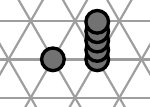}}\hspace{20pt}
		\subfloat[Configuration $\CC_2^2$]{\includegraphics[scale=1.5]{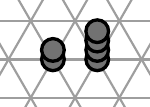}}\hspace{20pt}
        \subfloat[Configuration $\CC_3^2$]{\includegraphics[scale=1.5]{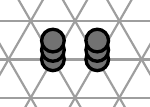}}
  \caption{Three configurations in $\Cset^2$}
  \label{fig:cset2}
\end{figure}

Figure \ref{fig:cset2} shows all the 2-point configurations ($\Cset^2$).
Note that all the robots in any of two configuration $\CC_2^2$ and $\CC_3^2$ have 
the same observation results due to the weak multiplicity.

\begin{figure}[tbhp]
  \centering
		\subfloat[$\View_2$]{\includegraphics[scale=1.3]{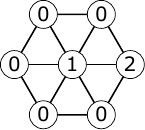}}\hspace{20pt}
		\subfloat[$\View_3$]{\includegraphics[scale=1.3]{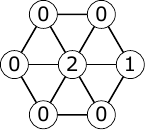}}\hspace{20pt}
        \subfloat[$\View_4$]{\includegraphics[scale=1.3]{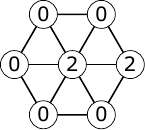}}
    \caption{Three views of the robots in $\Cset^2$}
  \label{fig:viewset2}
\end{figure}

Figure \ref{fig:viewset2} presents the view can be observed by the robots in 2-point configuration
(we call these three views $\View_2$, $\View_3$, and $\View_4$ respectively).
In any configuration in $\Cset^2$, 
each robot obtains one of the observation result among $\View_2$, $\View_3$ and $\View_4$.
By Lemma \ref{lem:nolocaldis},
considering any locally disconnecting rule is not required,
thus only the three rules depicted in Figure \ref{fig:ruleset2} have to be considered.

\begin{figure}[tbhp]
  \centering
		\subfloat[$\Rule_2$]{\includegraphics[scale=1.3]{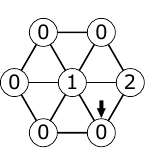}}\hspace{20pt}
		\subfloat[$\Rule_3$]{\includegraphics[scale=1.3]{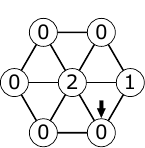}}\hspace{20pt}
        \subfloat[$\Rule_4$]{\includegraphics[scale=1.3]{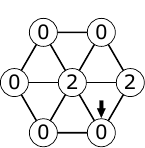}}
    \caption{Three feasible rules in $\Cset^2$}
  \label{fig:ruleset2}
\end{figure}

\begin{figure}[tbhp]
  \centering
		\includegraphics[scale=0.3]{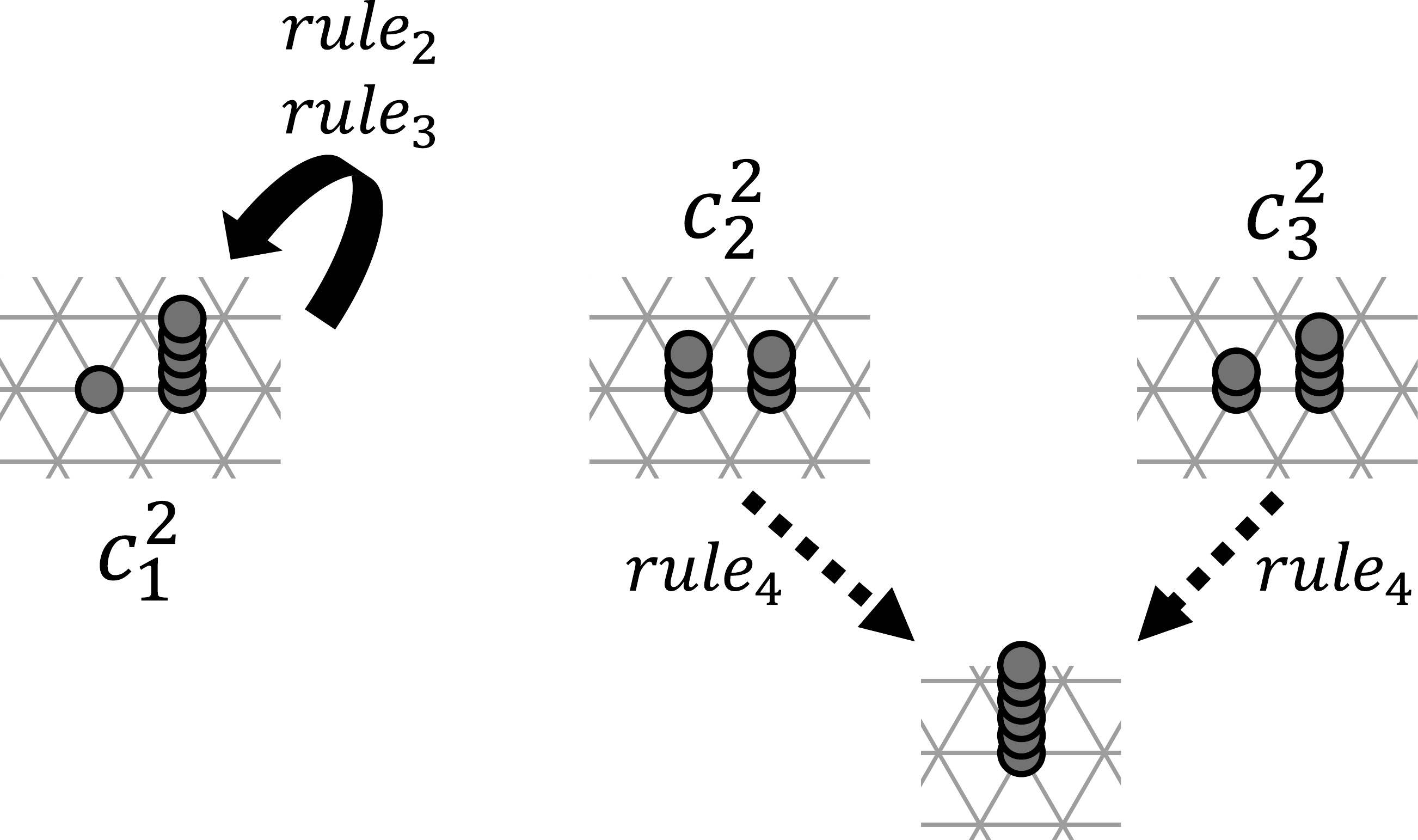}
  \caption{Configuration transition in $\Cset^2$}
  \label{fig:conftranset2}
\end{figure}

Assume that configuration $\CC_1^2$ is included in $\cycle$.
If algorithm $\Al$ includes $\Rule_2$ (resp. $\Rule_3$),
the scheduler activates a single robot (resp. all 5 robots located at the same point).
And the same configuration ($\CC_1^2$) appears again (refer to Figure \ref{fig:conftranset2}).
In the next activation, the scheduler can make the single robot (or the 5 robots)
to move back to the previously occupying point.
Thus configuration $\CC_1^2$ is not included in $\cycle$.

Now assume that configuration $\CC_2^2$ and/or $\CC_3^2$ is included in $\cycle$.
In this case, only $\View_4$ can be obtained by the robots.
If algorithm $\Al$ includes $\Rule_4$,
the scheduler can activate all the 6 robots, 
and make them to move to the same point causing 1-point configuration
by \emph{faulty concurrency}.
Therefore, configurations $\CC_2^2$ and $\CC_3^2$ are not included in $\cycle$.
\end{proof}

\begin{lemma}
\label{lem:noc3}
Cycle $\cycle$ never includes any configuration in $\Cset^3$.
\end{lemma}

\begin{proof}
First, we present all the configurations included in $\Cset^3$ as Figure \ref{fig:cset3}.
There are 15 3-point configurations, we denote the configurations by $\CC_1^3$ to $\CC_{15}^3$;
$\Cset^3 = \{\CC_1^3, \CC_2^3, \ldots, \CC_{15}^3\}$.

\begin{figure}[tbhp]
  \centering
		\subfloat[Configuration $\CC_1^3$]{\includegraphics[scale=1.3]{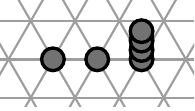}}\hspace{5pt}
		\subfloat[Configuration $\CC_2^3$]{\includegraphics[scale=1.3]{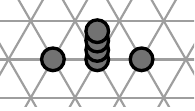}}\hspace{5pt}
        \subfloat[Configuration $\CC_3^3$]{\includegraphics[scale=1.3]{c3_3.pdf}}
        
		\subfloat[Configuration $\CC_4^3$]{\includegraphics[scale=1.3]{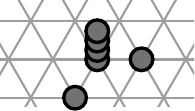}}\hspace{5pt}
		\subfloat[Configuration $\CC_5^3$]{\includegraphics[scale=1.3]{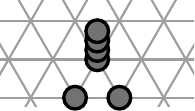}}\hspace{5pt}
        \subfloat[Configuration $\CC_6^3$]{\includegraphics[scale=1.3]{c3_6.pdf}}
        
		\subfloat[Configuration $\CC_7^3$]{\includegraphics[scale=1.3]{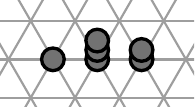}}\hspace{5pt}
		\subfloat[Configuration $\CC_8^3$]{\includegraphics[scale=1.3]{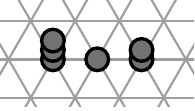}}\hspace{5pt}
        \subfloat[Configuration $\CC_9^3$]{\includegraphics[scale=1.3]{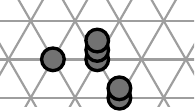}}
        
		\subfloat[Configuration $\CC_{10}^3$]{\includegraphics[scale=1.3]{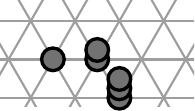}}\hspace{5pt}
		\subfloat[Configuration $\CC_{11}^3$]{\includegraphics[scale=1.3]{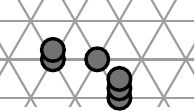}}\hspace{5pt}
		\subfloat[Configuration $\CC_{12}^3$]{\includegraphics[scale=1.3]{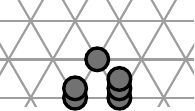}}
  
        \subfloat[Configuration $\CC_{13}^3$]{\includegraphics[scale=1.3]{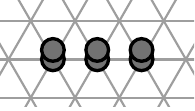}}\hspace{5pt}
		\subfloat[Configuration $\CC_{14}^3$]{\includegraphics[scale=1.3]{c3_14.pdf}}\hspace{5pt}
		\subfloat[Configuration $\CC_{15}^3$]{\includegraphics[scale=1.3]{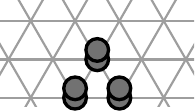}}
  \caption{All the configurations in $\Cset^3$}
  \label{fig:cset3}
\end{figure}

From configurations $\CC_1^3$ to $\CC_{15}^3$,
we can enumerate all the views can be obtained by any robots in $\Cset^3$. 
Figure \ref{fig:viewset2} shows all the views in any configuration in $\Cset^3$ (19 views exist).
Note that views $\View_2$, $\View_3$, and $\View_4$ are the same as them appeared in $\Cset^2$.

\begin{figure}[tbhp]
  \centering
		\subfloat[$\View_2$]{\includegraphics[scale=1]{view2.pdf}}\hspace{8pt}
		\subfloat[$\View_3$]{\includegraphics[scale=1]{view3.pdf}}\hspace{8pt}
        \subfloat[$\View_4$]{\includegraphics[scale=1]{view4.pdf}}\hspace{8pt}
		\subfloat[$\View_5$]{\includegraphics[scale=1]{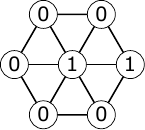}}\hspace{8pt}
        \subfloat[$\View_6$]{\includegraphics[scale=1]{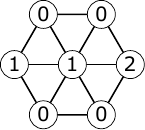}}
        
        \subfloat[$\View_7$]{\includegraphics[scale=1]{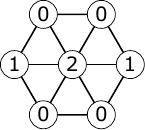}}\hspace{8pt}
        \subfloat[$\View_8$]{\includegraphics[scale=1]{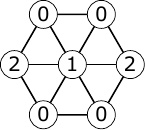}}\hspace{8pt}
		\subfloat[$\View_9$]{\includegraphics[scale=1]{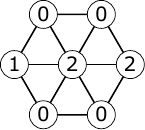}}\hspace{8pt}
        \subfloat[$\View_{10}$]{\includegraphics[scale=1]{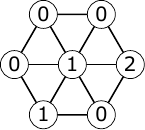}}\hspace{8pt}
		\subfloat[$\View_{11}$]{\includegraphics[scale=1]{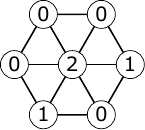}}

        \subfloat[$\View_{12}$]{\includegraphics[scale=1]{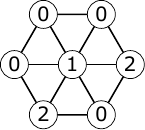}}\hspace{8pt}
        \subfloat[$\View_{13}$]{\includegraphics[scale=1]{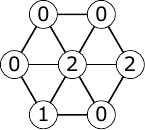}}\hspace{8pt}
		\subfloat[$\View_{14}$]{\includegraphics[scale=1]{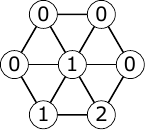}}\hspace{8pt}
		\subfloat[$\View_{15}$]{\includegraphics[scale=1]{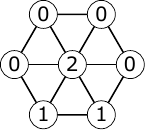}}\hspace{8pt}
        \subfloat[$\View_{16}$]{\includegraphics[scale=1]{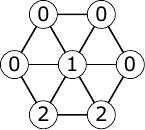}}

 	\subfloat[$\View_{17}$]{\includegraphics[scale=1]{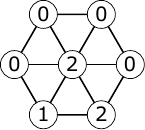}}\hspace{8pt}
		\subfloat[$\View_{18}$]{\includegraphics[scale=1]{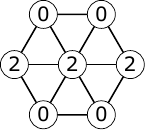}}\hspace{8pt}
        \subfloat[$\View_{19}$]{\includegraphics[scale=1]{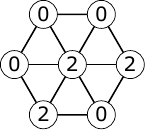}}\hspace{8pt}
        \subfloat[$\View_{20}$]{\includegraphics[scale=1]{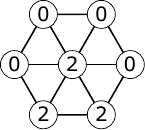}}
    \caption{All views can be obtained by a robot in $\Cset^3$}
  \label{fig:viewset3}
\end{figure}

From all the views obtained by a robot 
in any configuration in $\Cset^3$,
we can enumerate all feasible rules.
Also in this case, 
we do not consider a locally disconnecting rule (due to Lemma \ref{lem:nolocaldis}).
Moreover, we also exclude a rule that makes a robot to any point occupied by another robot 
because such a rule causes a configuration in $\Cset^1$ or $\Cset^2$ 
(refer to Corollary \ref{cor:samesetk}).
As a result, we obtain the 16 rules depicted in Figure \ref{fig:ruleset3}.

\begin{figure}[tbhp]
  \centering
		\subfloat[$\Rule_2$]{\includegraphics[scale=1]{rule2g.pdf}}\hspace{20pt}
		\subfloat[$\Rule_3$]{\includegraphics[scale=1]{rule3g.pdf}}\hspace{20pt}
        \subfloat[$\Rule_4$]{\includegraphics[scale=1]{rule4g.pdf}}\hspace{20pt}
		\subfloat[$\Rule_5$]{\includegraphics[scale=1]{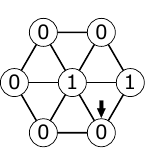}}

        \subfloat[$\Rule_6$]{\includegraphics[scale=1]{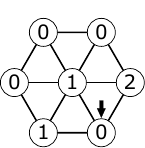}}\hspace{20pt}
        \subfloat[$\Rule_7$]{\includegraphics[scale=1]{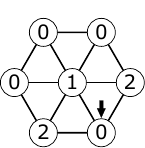}}\hspace{20pt}
        \subfloat[$\Rule_8$]{\includegraphics[scale=1]{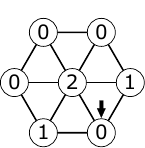}}\hspace{20pt}
		\subfloat[$\Rule_9$]{\includegraphics[scale=1]{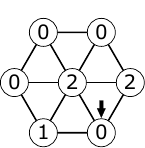}}
  
        \subfloat[$\Rule_{10}$]{\includegraphics[scale=1]{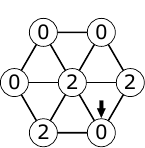}}\hspace{20pt}
		\subfloat[$\Rule_{11}$]{\includegraphics[scale=1]{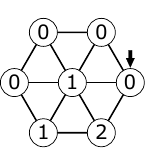}}\hspace{20pt}
        \subfloat[$\Rule_{12}$]{\includegraphics[scale=1]{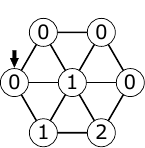}}\hspace{20pt}
        \subfloat[$\Rule_{13}$]{\includegraphics[scale=1]{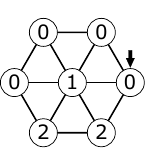}}
        
		\subfloat[$\Rule_{14}$]{\includegraphics[scale=1]{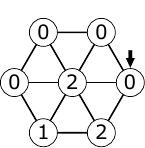}}\hspace{20pt}
		\subfloat[$\Rule_{15}$]{\includegraphics[scale=1]{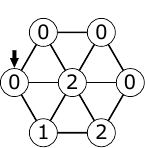}}\hspace{20pt}
        \subfloat[$\Rule_{16}$]{\includegraphics[scale=1]{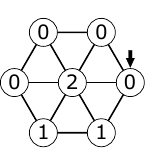}}\hspace{20pt}
 	\subfloat[$\Rule_{17}$]{\includegraphics[scale=1]{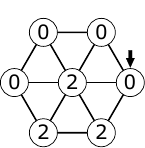}}
  \caption{All feasible rules in $\Cset^3$}
  \label{fig:ruleset3}
\end{figure}

Now we can obtain the induced subgraph $\overrightarrow{G_{Al}}(\Cset^3)$ 
of $\overrightarrow{G_{Al}}$ 
because the set of arcs in $\overrightarrow{G_{Al}}(\Cset^3)$ is defined by 
the rules in Figure \ref{fig:ruleset3}.
It is worthwhile to mention that 
no robot can distinguish the difference between $\CC_6^3$ and $\CC_7^3$,
and between $\CC_9^3$ and $\CC_{10}^3$ due to the weak multiplicity.

\begin{figure}[tbhp]
  \centering
		\includegraphics[scale=0.15]{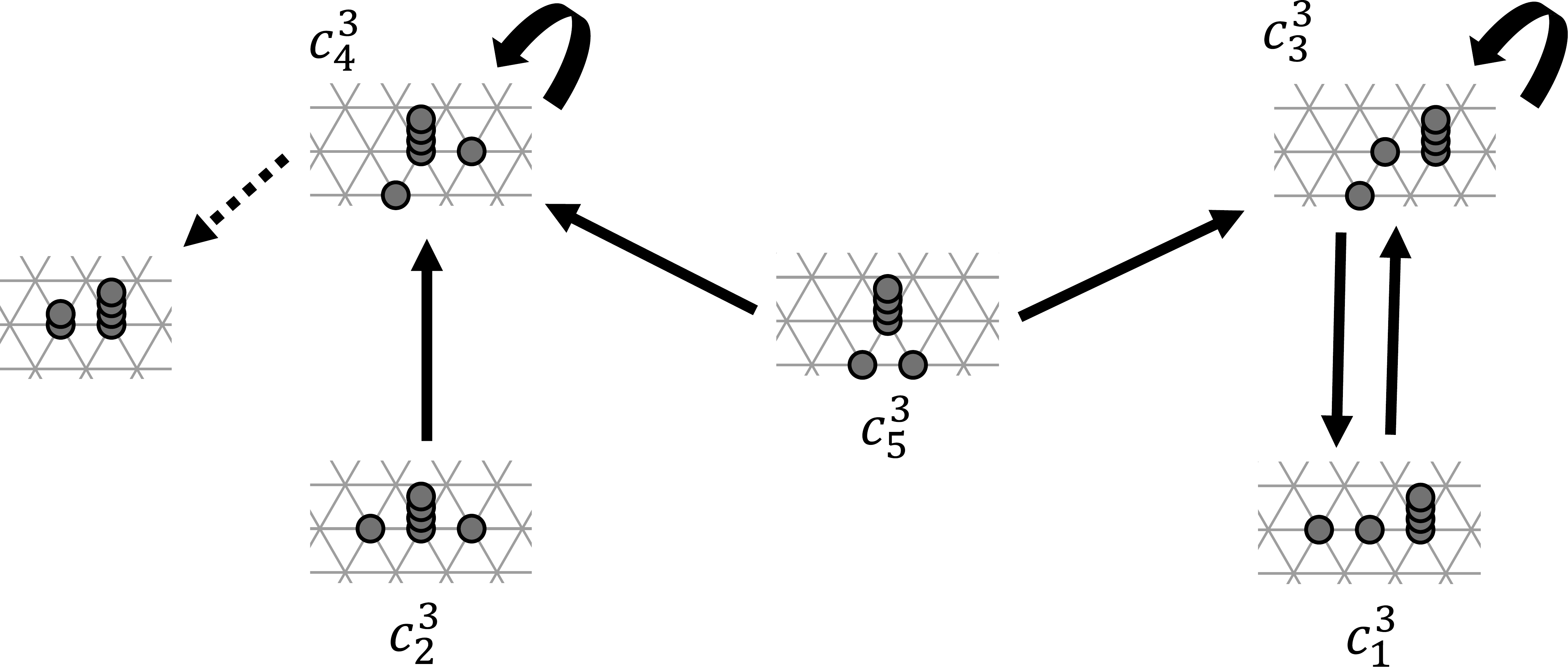}

        \vspace{20pt}
 	\includegraphics[scale=0.15]{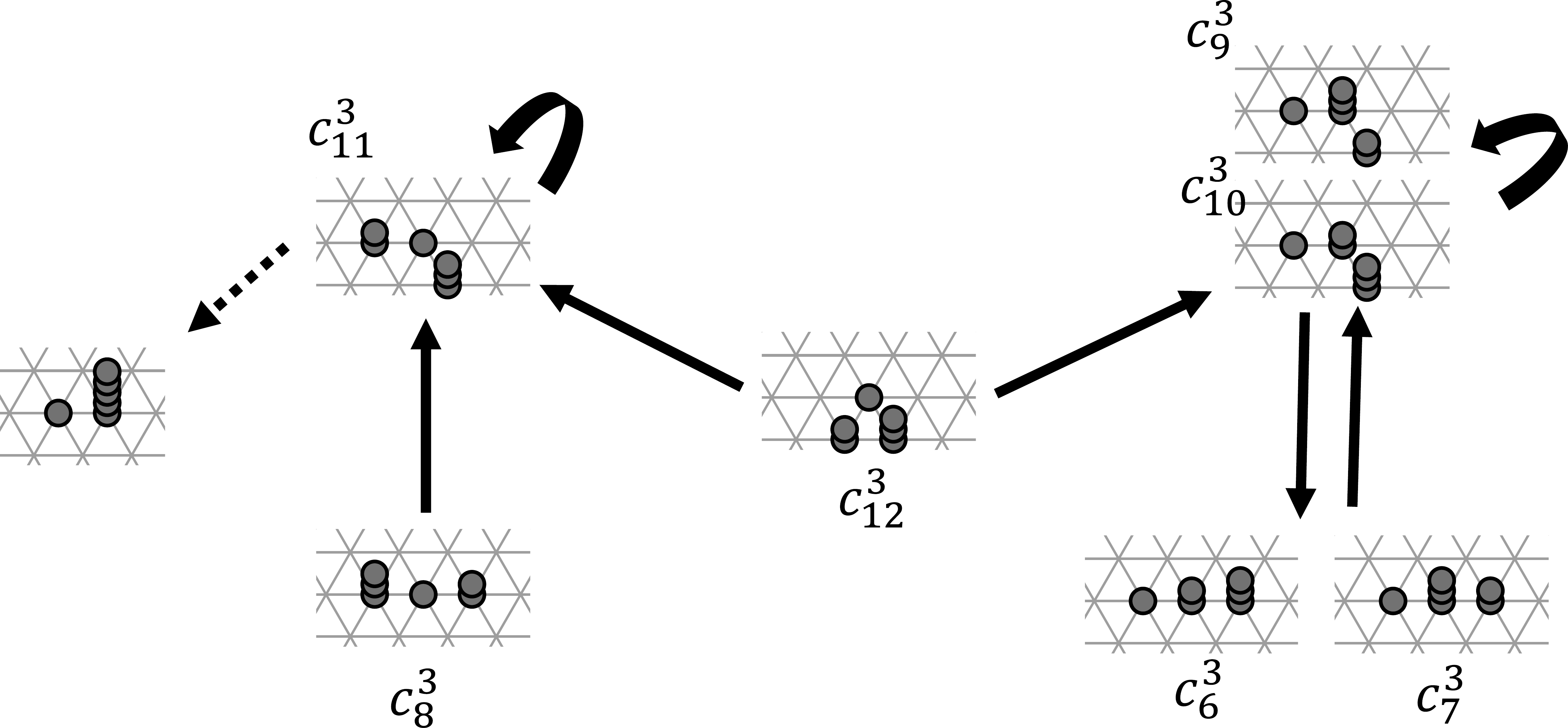}

        \vspace{20pt}
		\includegraphics[scale=0.15]{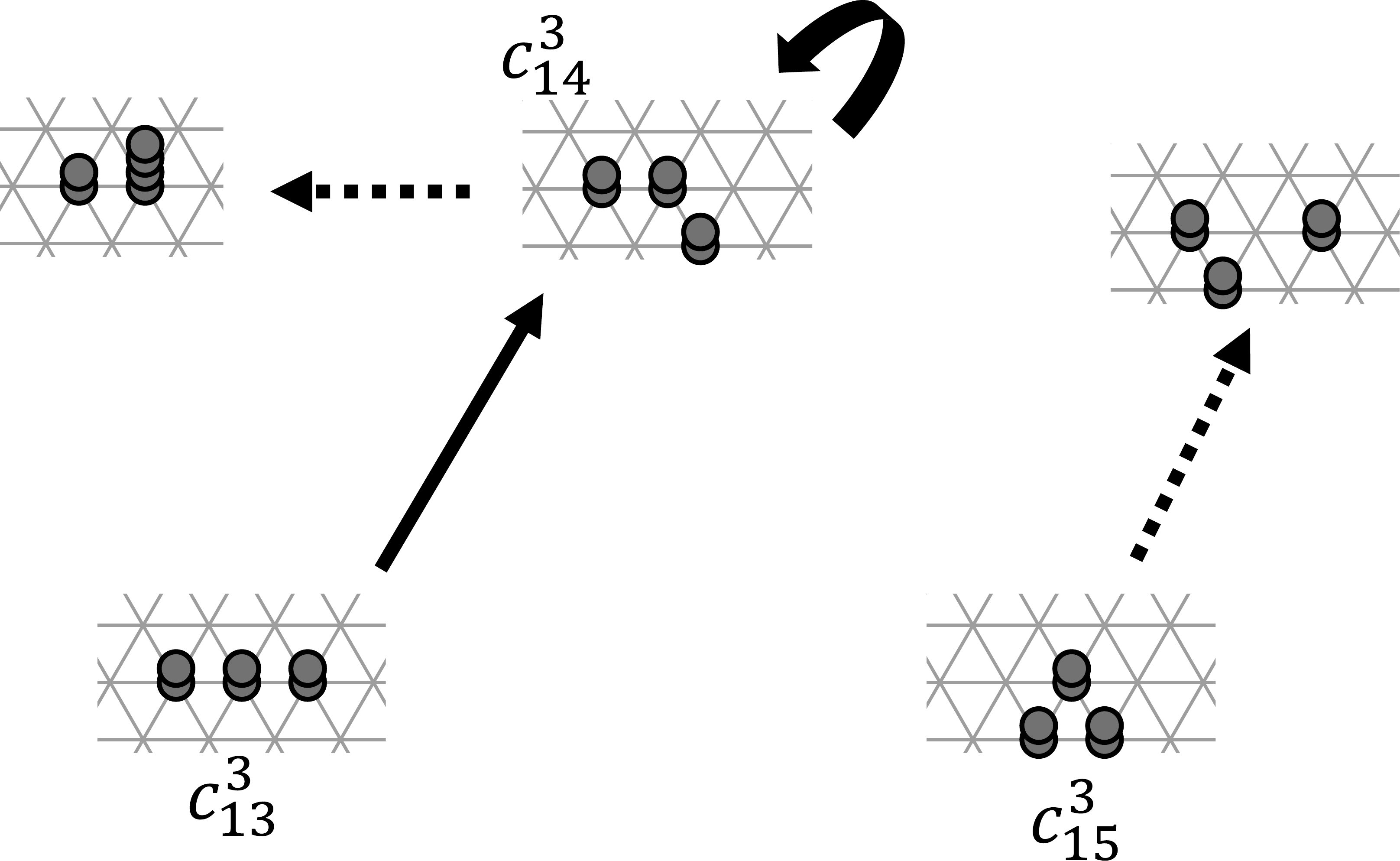}
\caption{Configuration transition in $\Cset^3$ (\ie the induced subgraph $\overrightarrow{G_{Al}}(\Cset^3)$ of $\overrightarrow{G_{Al}}$)}
  \label{fig:conftranset3}
\end{figure}

Figure \ref{fig:conftranset3} presents the induced subgraph $\overrightarrow{G_{Al}}(\Cset^3)$.
A dotted arc represents an execution occurring a disconnected configuration or 
a smaller $k$-point configuration due to a faulty concurrency.
There are 6 (directed) loops 
at configurations $\CC_3^3$, $\CC_4^3$, $\CC_9^3$, $\CC_{10}^3$, $\CC_{11}^3$, and $\CC_{14}^3$.
An infinity execution of these loops are the same;
only the robots located at the center point repeatedly move back and forth between two points.
Thus these loops cannot become cycle $\cycle$.

Now we consider the following two (directed) cycles,
($\CC_1^3 \to \CC_3^3 \to \CC_1^3$) and
($\CC_6^3 \to \CC_{10}^3 \to \CC_6^3$).
and ($\CC_7^3 \to \CC_9^3 \to \CC_7^3$).
However in any cycles,
only some robots at the same point
infinitely move back and forth between two adjacent points.
This implies that there exists a robot which never moves,
hence, these cycles cannot become cycle $\cycle$.
\end{proof}

We can prove the remaining parts of the proof below using the same manner, 
but each proof becomes much more longer, 
thus we omit the detailed proofs of the following three lemmas for $\Cset^4$, $\Cset^5$, and $\Cset^6$ respectively here.

\begin{lemma}
\label{lem:noc4}
Cycle $\cycle$ never includes any configuration in $\Cset^4$.
\end{lemma}

\begin{figure}[tbhp]
  \centering
		\includegraphics[scale=0.125]{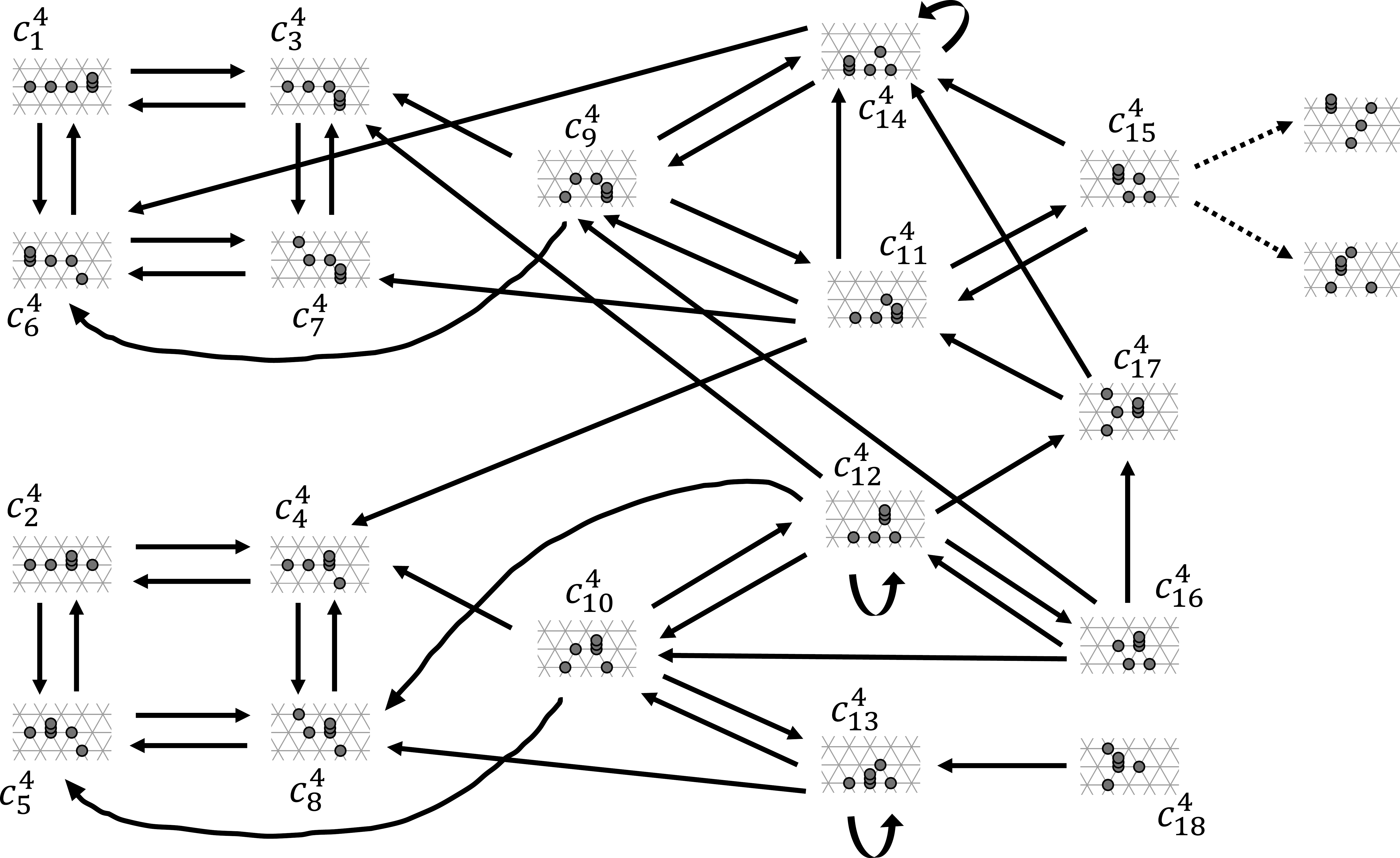}
    \caption{Configuration transition in $\Cset^4$ (\ie the induced subgraph $\overrightarrow{G_{Al}}(\Cset^4)$ of $\overrightarrow{G_{Al}}$): From $\CC_1^4$ to $\CC_{18}^4$}
  \label{fig:conftranset4part1}
\end{figure}

\begin{sidewaysfigure}[p]
  \centering
 	\includegraphics[scale=0.18]{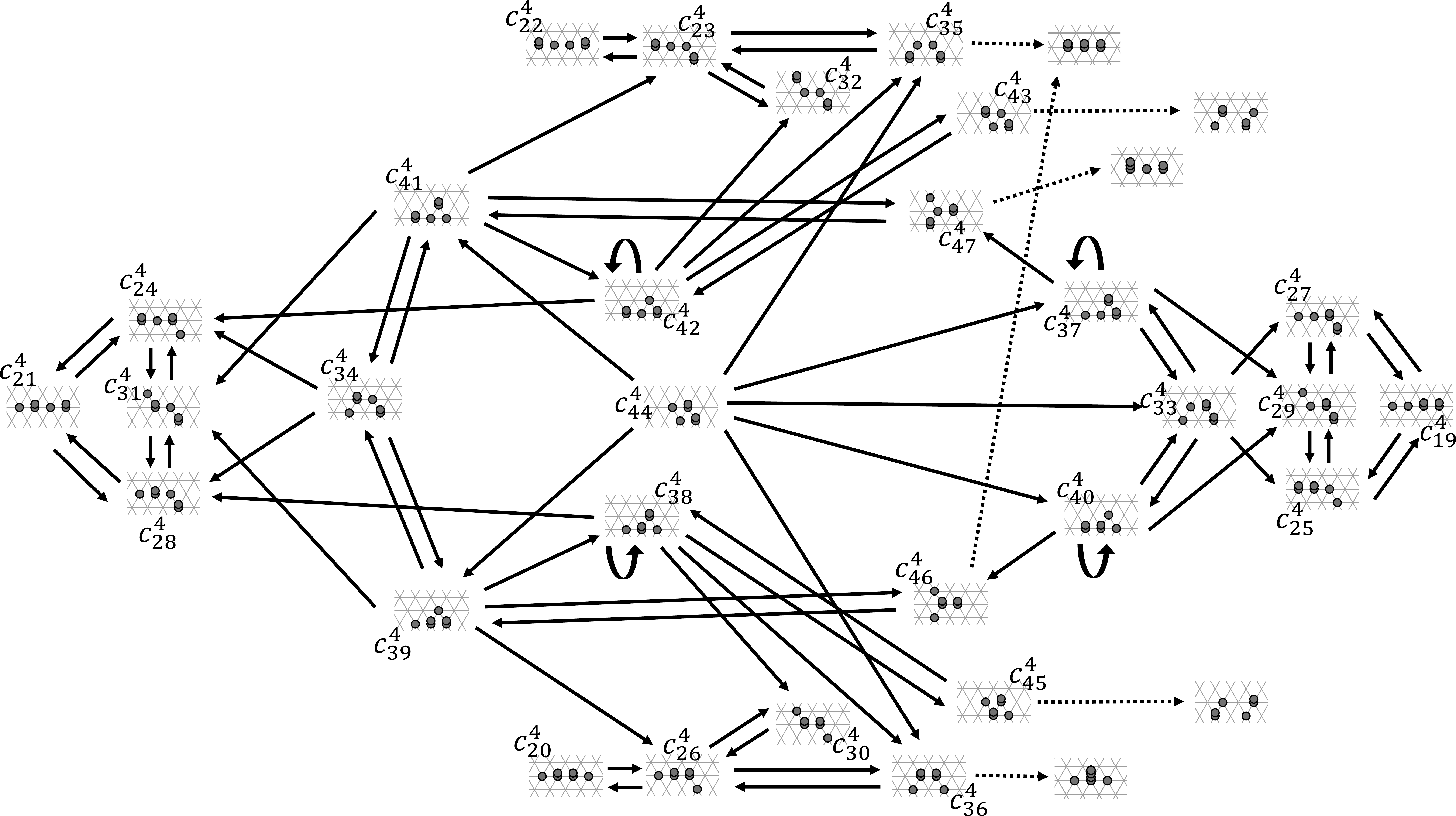}
    \caption{Configuration transition in $\Cset^4$ (\ie the induced subgraph $\overrightarrow{G_{Al}}(\Cset^4)$ of $\overrightarrow{G_{Al}}$): From $\CC_{19}^4$ to $\CC_{47}^4$}
  \label{fig:conftranset4part2}
\end{sidewaysfigure}

\begin{proof}
(Sketch)
There are 47 configurations in $\Cset^4$, 
and 40 views can be obtained by a robot in any configuration in $\Cset^4$.
From these views, we can obtain 36 feasible rules in $\Cset^4$.
A resultant induced subgraph $\overrightarrow{G_{Al}}(\Cset^4)$ of $\overrightarrow{G_{Al}}$)
is given in Figures \ref{fig:conftranset4part1} and \ref{fig:conftranset4part2}.
There are many cycles in $\overrightarrow{G_{Al}}(\Cset^4)$,
however, no appropriate cycle in $\overrightarrow{G_{Al}}(\Cset^4)$ which can become feasible $\cycle$.
\end{proof}

\newpage

\begin{lemma}
\label{lem:noc5}
Cycle $\cycle$ never includes any configuration in $\Cset^5$.
\end{lemma}

\begin{lemma}
\label{lem:noc6}
Cycle $\cycle$ never includes any configuration in $\Cset^6$.
\end{lemma}

We omit the proof of Lemmas \ref{lem:noc5} and \ref{lem:noc6}
however these are can be proven by the same manner;
there exist 87 and 82 configurations in $\overrightarrow{G_{Al}}(\Cset^5)$ and $\overrightarrow{G_{Al}}(\Cset^6)$ respectively.

From lemmas \ref{lem:noc1}, \ref{lem:noc2}, \ref{lem:noc3}, \ref{lem:noc4},
\ref{lem:noc5}, and \ref{lem:noc6},
the following lemma holds.

\begin{lemma}
\label{lem:nocycle}
    There is no directed cycle $\cycle$ in $\overrightarrow{G_\Al} = (G, A_\Al)$
    implying the valid infinite execution of the perpetual marching.
\end{lemma}

Therefore, the following theorem holds by Lemma \ref{lem:nocycle}.

\begin{theorem}
\label{thm:impos6}
Under an $\SSYNC$ scheduler, no deterministic algorithm exists that can solve the perpetual marching problem using unoriented (\emph{i.e.,} lacking geometric agreement) LCM-robots with a visibility range of 1, provided the number of robots is six.
\end{theorem}

Theorem \ref{thm:impos6} presents an impossibility result only for 6 LCM-robots,
however, an impossibility result for 5 or less LCM-robots can be easily proven.
In particular, the impossibility results for 1, 2, or 3 LCM-robots
can be proven by a straight-forward manner (only the small number of all connected configurations
exist),
and the results for 4 or 5 LCM-robots can be proven as the same manner as the case for 6 LCM-robots.
However, many of $k$-point configurations appeared when the number of robots is 4 or 5 
are also appeared in the proof for 6 LCM-robots, 
thus, there are only some configurations have to be newly considered to prove the impossibility result.
As a result, Theorem \ref{thm:impos} holds. 

\newpage
\section{The 7-pairbots-gathering Problem}
\label{sec:7gather}

\subsection{Problem Definition}
In this section, we consider the 7-pairbots-gathering problem,
which is the gathering problem of 7 \PR s on a triangular grid.
Generally, many gathering problems aim to gather all robots to one common point, 
but here we consider the gathering problem 
where two or more robots do not exist at the same point 
like the gathering considered in fat robot models \cite{fatrobot}.
This means that our goal is to gather the robots as close together as possible.

We define the 7-\PR s-gathering problem as the following.

\begin{definition}
\emph{\bf{7-pairbots-gathering Problem.}} 
Given an arbitrary connected configuration consisting of 7 pairbots in short state,
algorithm $\Al$ solves the 7-pairbots-gathering problem 
if $\Al$ satisfies all the following conditions:
(1) algorithm $\Al$ eventually terminates;
algorithm $\Al$ eventually reaches a configuration such that no robot can move,
(2) all pairbots are in short state and no two \PR s exist at the same point when algorithm $\Al$ terminates, 
and (3) $dist(r_i, r_j) \leq 2$ for any non-paired two robots $r_i$ and $r_j$ when algorithm $\Al$ terminates.
\end{definition}

\begin{figure}[tb]
  \centering
   \includegraphics[scale=4]{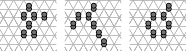}
  \caption{Examples of initial configurations}
  \label{initex}
\end{figure}

\begin{figure}[tb]
  \centering
   \includegraphics[scale=4]{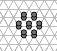}
  \caption{Goal configuration}
  \label{gathering}
\end{figure}

Figure \ref{initex} shows three examples of initial configurations 
and Figure \ref{gathering} illustrates the goal configuration of 7-pairbots-gathering problem.
Note that only one configuration as Figure \ref{gathering} is allowed as a goal configuration.

We can also consider the \emph{7-robots-gathering problem}
which is the gathering by 7 LCM-robots (robots in \emph{LCM} model) instead of \PR s.
This problem is already introduced in \cite{shibata2022visibility},
and the two following theorems hold by the literature.

\begin{theorem}\emph{\cite{shibata2022visibility}}
\label{lem:unsol1}
    For robots with visibility range 1,
    there exists no deterministic algorithm to solve the 7-robots-gathering problem 
    even under an $\FSYNC$ scheduler.
\end{theorem}

\begin{theorem}\emph{\cite{shibata2022visibility}}\label{lem:sol1}
    For robots with visibility range 2,
    there exists a deterministic algorithm to solve the 7-robots-gathering problem 
    from any connected initial configuration under an $\FSYNC$ scheduler.
\end{theorem}

Theorems~\ref{lem:unsol1} and~\ref{lem:sol1} suggest that the visibility range plays a critical role in the solvability of the 7-robots-gathering problem, especially under deterministic algorithms and an $\FSYNC$ scheduler. While it is unsolvable for robots with a visibility range of 1, the problem becomes solvable for robots with a visibility range of 2 if they start from any connected initial configuration.
In the following, we show that 
the 7-pairbots-gathering problem 
for \PR s is solvable
even with visibility range 1.

\subsection{7-pairbots-gathering Algorithm}

Here we introduce the proposed algorithm to solve the 7-pairbots-gathering algorithm
using graphical representation.
The detailed pseudocode of the algorithm will be presented in the next section.

\begin{figure*}[p]
  \centering
		\includegraphics[scale=2.2]{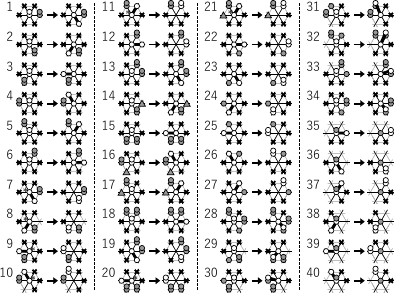}
     \includegraphics[scale=2.2]{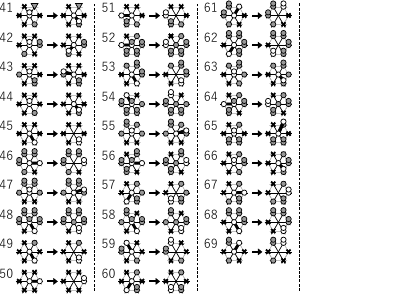}
  \caption{The proposed algorithm for the 7-pairbots-gathering}
  \label{fig:7pg-graphic}
\end{figure*}

Figures \ref{fig:7pg-graphic}
shows the proposed algorithm
by \PR s with a common sense of directions.
Each white circle shows a \PR~that executes the proposed algorithm, 
and each black circle represents the observed robots.
Each cross mark means a point which is not occupied by any robot, 
a triangle presents a point occupied one or more robots (in rules 14, 16, 17, and 21), 
and an inverted (reversed) triangle represents a point occupied two robots or no robot (in rule 41).
Some rules for a \PR~in \LN~state 
marks a asterisk near the \PR,
which means the robot executes the rule even if there are some other robots at the same (\ie center) point
(in rules from 7 to 11, 20, and 21).
If there is no any mark at the point, 
the algorithm does not care about the point.

\subsection{Pseudocode of the 7-pairbots-gathering Algorithm}
\label{sec:apb}
In this section, we present the pseudocode of the proposed algorithm 
to solve the 7-pairbots-gathering problem.
We assume that all robots agree on the orientation and directions of both axes,
which is the same assumption as the literature \cite{shibata2022visibility};
every robot labels the adjacent point located on its right side as $\ell_1$, 
and then labels the other adjacent points from $\ell_2$ to $\ell_6$ in clockwise order
(refer to the labeling of robot $r_i$ in Figure \ref{fig:label}).
The point where a robot currently exists is labeled by $\ell_0$.
Algorithm \ref{alg:7pg} shows the proposed algorithm;
as the same as the algorithm in the previous section, 
each rule is presented as \textsf{[Rule No.]: [Condition] $\to$ [Action]}.

\begin{algorithm}[thbp]
    \caption{Algorithm for 7-pairbots-gathering}
    \label{alg:7pg}
    {\bf variables and functions:}
        \begin{description}
        \item[$~\cdot~ \bud \in \{\ell_0, \ell_1, \ldots, \ell_6\}$~:] A label where its \bd~exists.
        \item[$~\cdot~ \chk(\ell_x, \nr)$~:] A function for checking the number of robots
        that returns $\mathsf{TRUE}$ when 
        the number of robots on the adjacent point at $\ell_x$ is the same as $\nr$. 
        Note that $\nr \in \{0,1,2\}$ due to weak multiplicity detection.
        The first parameter can be a set of labels $\mathbf{L} \subseteq \{\ell_x | 0 \leq x \leq 6 \}$; 
        the function returns $\mathsf{TRUE}$ only when $\chk(\forall \ell_x \in \mathbf{L}, \nr) = \mathsf{TRUE}$.
        \item[$~\cdot~ \move(\ell_x)$~:] Move to $l_x$.
\end{description}
\vspace{3pt}
{\bf algorithm:}
    \begin{algorithmic}
\State Rule1: $\bud=l_0\land \chk(\{l_2,l_3,l_4,l_5,l_6\},0) \land \chk(l_1,2) \to \move(l_2)$
\State Rule2: $\bud=l_0\land \chk(\{l_1,l_3,l_4,l_5,l_6\},0) \land \chk(l_2,2) \to \move(l_3)$
\State Rule3: $\bud=l_0\land \chk(\{l_1,l_2,l_4,l_5,l_6\},0) \land \chk(l_3,2) \to \move(l_4)$
\State Rule4: $\bud=l_0\land \chk(\{l_1,l_2,l_3,l_5,l_6\},0) \land \chk(l_4,2) \to \move(l_5)$
\State Rule5: $\bud=l_0\land \chk(\{l_1,l_2,l_3,l_4,l_6\},0) \land \chk(l_5,2) \to \move(l_6)$
\State Rule6: $\bud=l_0\land \chk(\{l_1,l_2,l_3,l_4,l_5\},0) \land \chk(l_6,2) \to \move(l_1)$
\State Rule7: $\bud=l_2\land \chk(\{l_3,l_4,l_5,l_6\},0) \land \chk(l_2,1) \land \chk(l_1,2) \to \move(l_2)$
\State Rule8: $\bud=l_3\land \chk(\{l_1,l_4,l_5,l_6\},0) \land \chk(l_3,1) \land \chk(l_2,2) \to \move(l_3)$
\State Rule9: $\bud=l_4\land \chk(\{l_1,l_2,l_5,l_6\},0) \land \chk(l_4,1) \land \chk(l_3,2) \to \move(l_4)$
\State Rule10: $\bud=l_5\land \chk(\{l_1,,l_2,l_6\},0) \land \chk(l_5,1) \land \chk(l_4,2) \to \move(l_5)$
\State Rule11: $\bud=l_6\land \chk(\{l_1,l_2,l_3,l_4\},0) \land \chk(l_6,1) \land \chk(l_5,2) \to \move(l_6)$
\State Rule12: $\bud=l_1\land \chk(\{l_2,l_3,l_4,l_5\},0) \land \chk(\{l_0,l_1\},1) \land \chk(l_6,2) \to \move(l_1)$
\State Rule13: $\bud=l_0\land \chk(\{l_2,l_3,l_4,l_5\},0) \land \chk(\{l_1,l_6\},2) \to \move(l_2)$
\State Rule14: $\bud=l_0\land \chk(\{l_3,l_4,l_5,l_6\},0) \land \neg\chk(l_1,0) \land \chk(l_2,2) \to \move(l_3)$
\State Rule15: $\bud=l_0\land \chk(\{l_1,l_4,l_5,l_6\},0) \land \chk(\{l_2,l_3\},2) \to \move(l_4)$
\State Rule16: $\bud=l_0\land \chk(\{l_1,l_2,l_5,l_6\},0) \land \neg\chk(l_3,0) \land \chk(l_4,2) \to \move(l_5)$
\State Rule17: $\bud=l_0\land \chk(\{l_1,l_2,l_3,l_6\},0) \land \neg\chk(l_4,0) \land \chk(l_5,2) \to \move(l_6)$
\State Rule18: $\bud=l_0\land \chk(\{l_1,l_2,l_3,l_4\},0) \land \chk(\{l_5,l_6\},2) \to \move(l_1)$
\State Rule19: $\bud=l_2\land \chk(\{l_3,l_4,l_5\},0) \land \chk(\{l_0,l_2\},1) \land \chk(\{l_1,l_6\},2) \to \move(l_2)$
\State Rule20: $\bud=l_4\land \chk(\{l_1,l_5,l_6\},0) \land \chk(l_4,1) \land \chk(\{l_2,l_3\},2) \to \move(l_4)$
\State Rule21: $\bud=l_6\land \chk(\{l_1,l_2,l_3\},0) \land \neg\chk(l_4,0) \land \chk(l_5,2) \land \chk(l_6,1) \to \move(l_6)$
\State Rule22: $\bud=l_1\land \chk(\{l_3,l_4,l_5,l_6\},0) \land \chk(\{l_0,l_1,l_2\},1) \to \move(l_1)$
\State Rule23: $\bud=l_2\land \chk(\{l_1,l_4,l_5,l_6\},0) \land \chk(\{l_0,l_2,l_3\},1) \to \move(l_2)$
\State Rule24: $\bud=l_3\land \chk(\{l_1,l_2,l_5,l_6\},0) \land \chk(\{l_0,l_3,l_4\},1) \to \move(l_3)$
\State Rule25: $\bud=l_4\land \chk(\{l_1,l_2,l_3,l_6\},0) \land \chk(\{l_0,l_4,l_5\},1) \to \move(l_4)$
\end{algorithmic}
\end{algorithm}

\begin{algorithm}[thbp]
  \begin{algorithmic}
\State Rule26: $\bud=l_5\land \chk(\{l_1,l_2,l_3,l_4\},0) \land \chk(\{l_0,l_5,l_6\},1) \to \move(l_5)$
\State Rule27: $\bud=l_6\land \chk(\{l_2,l_3,l_4,l_5\},0) \land \chk(\{l_0,l_1,l_6\},1) \to \move(l_6)$
\State Rule28: $\bud=l_0\land \chk(\{l_3,l_4,l_6\},0) \land \chk(l_2,1) \land \chk(\{l_1,l_5\},2) \to \move(l_2)$
\State Rule29: $\bud=l_0\land \chk(\{l_4,l_5,l_6\},0) \land \chk(l_3,1) \land \chk(l_2,2) \to \move(l_3)$
\State Rule30: $\bud=l_0\land \chk(\{l_1,l_5,l_6\},0) \land \chk(l_4,1) \land \chk(l_3,2) \to \move(l_4)$
\State Rule31: $\bud=l_0\land \chk(\{l_1,l_2,l_6\},0) \land \chk(l_5,1) \land \chk(l_4,2) \to \move(l_5)$
\State Rule32: $\bud=l_0\land \chk(\{l_1,l_2,l_3\},0) \land \chk(l_6,1) \land \chk(l_5,2) \to \move(l_6)$
\State Rule33: $\bud=l_0\land \chk(\{l_2,l_3,l_4\},0) \land \chk(l_1,1) \land \chk(l_6,2) \to \move(l_1)$
\State Rule34: $\bud=l_0\land \chk(\{l_3,l_4,l_5,l_6\},0) \land \chk(l_2,1) \land \chk(l_1,2) \to \move(l_2)$
\State Rule35: $\bud=l_1\land \chk(\{l_2,l_3\},0) \land \chk(l_1,1) \land \chk(l_0,2) \to \move(l_1)$
\State Rule36: $\bud=l_2\land \chk(\{l_3,l_4\},0) \land \chk(l_2,1) \land \chk(l_0,2) \to \move(l_2)$
\State Rule37: $\bud=l_6\land \chk(\{l_1,l_2\},0) \land \chk(l_6,1) \land \chk(l_0,2) \to \move(l_6)$
\State Rule38: $\bud=l_3\land \chk(\{l_4,l_5\},0) \land \chk(l_3,1) \land \chk(l_0,2) \to \move(l_3)$
\State Rule39: $\bud=l_4\land \chk(\{l_5,l_6\},0) \land \chk(l_4,1) \land \chk(l_0,2) \to \move(l_4)$
\State Rule40: $\bud=l_5\land \chk(\{l_1,l_6\},0) \land \chk(l_5,1) \land \chk(l_0,2) \to \move(l_5)$
\State Rule41: $\bud=l_0\land \chk(\{l_1,l_3,l_4,l_5\},0) \land \chk(l_2,1) \land \neg\chk(l_6,1) \to \move(l_2)$
\State Rule42: $\bud=l_0\land \chk(\{l_2,l_4,l_5,l_6\},0) \land \chk(l_3,1) \land \chk(l_1,2) \to \move(l_3)$
\State Rule43: $\bud=l_0\land \chk(\{l_1,l_3,l_5,l_6\},0) \land \chk(l_4,1) \land \chk(l_2,2) \to \move(l_4)$
\State Rule44: $\bud=l_0\land \chk(\{l_1,l_3,l_4,l_5,l_6\},0) \land \chk(l_2,1) \to \move(l_2)$
\State Rule45: $\bud=l_2\land \chk(\{l_1,l_3,l_4,l_5,l_6\},0) \land \chk(\{l_0,l_2\},1) \to \move(l_2)$
\State Rule46: $\bud=l_1\land \chk(l_2,0) \land \chk(\{l_0,l_1,l_3\},1) \land \chk(\{l_4,l_5,l_6\},2) \to \move(l_1)$
\State Rule47: $\bud=l_0\land \chk(\{l_2,l_3\},0) \land \chk(\{l_1,l_4\},1) \land \chk(\{l_5,l_6\},2) \to \move(l_1)$
\State Rule48: $\bud=l_2\land \chk(l_3,0) \land \chk(l_2,1) \land \chk(\{l_0,l_1,l_4,l_5,l_6\},2) \to \move(l_2)$
\State Rule49: $\bud=l_2\land \chk(\{l_1,l_3,l_4,l_5\},0) \land \chk(\{l_0,l_2,l_6\},1) \to \move(l_2)$
\State Rule50: $\bud=l_1\land \chk(\{l_2,l_3,l_4,l_5,l_6\},0) \land \chk(\{l_0,l_1\},1) \to \move(l_1)$
\State Rule51: $\bud=l_4\land \chk(\{l_1,l_5,l_6\},0) \land \chk(\{l_0,l_2,l_4\},1) \land \chk(l_3,2) \to \move(l_4)$
\State Rule52: $\bud=l_4\land \chk(l_5,0) \land \chk(\{l_4,l_6\},1) \land \chk(\{l_0,l_1,l_2,l_3\},2) \to \move(l_4)$
\State Rule53: $\bud=l_2\land \chk(\{l_3,l_4\},0) \land \chk(\{l_0,l_2,l_5\},1) \land \chk(\{l_1,l_6\},2) \to \move(l_2)$
\State Rule54: $\bud=l_5\land \chk(l_6,0) \land \chk(\{l_1,l_5\},1) \land \chk(\{l_0,l_2,l_3,l_4\},2) \to \move(l_5)$
\State Rule55: $\bud=l_0\land \chk(\{l_2,l_3\},0) \land \chk(\{l_1,l_4,l_6\},1) \land \chk(l_5,2) \to \move(l_1)$
\State Rule56: $\bud=l_1\land \chk(\{l_2,l_5\},0) \land \chk(l_1,1) \land \chk(\{l_0,l_3,l_4,l_6\},2) \to \move(l_1)$
\State Rule57: $\bud=l_3\land \chk(\{l_4,l_5\},0) \land \chk(\{l_0,l_1,l_3,l_6\},1) \land \chk(l_2,2) \to \move(l_3)$
\State Rule58: $\bud=l_2\land \chk(\{l_3,l_6\},0) \land \chk(\{l_2,l_4\},1) \land \chk(\{l_0,l_1,l_5\},2) \to \move(l_2)$
\State Rule59: $\bud=l_5\land \chk(\{l_1,l_3,l_6\},0) \land \chk(\{l_0,l_2,l_5\},1) \land \chk(l_4,2) \to \move(l_5)$
\State Rule60: $\bud=l_3\land \chk(\{l_1,l_4,l_5\},0) \land \chk(\{l_0,l_3,l_6\},1) \land \chk(l_2,2) \to \move(l_3)$
\State Rule61: $\bud=l_6\land \chk(\{l_1,l_2\},0) \land \chk(\{l_0,l_3,l_6\},1) \land \chk(\{l_4,l_5,2) \to \move(l_6)$
\State Rule62: $\bud=l_3\land \chk(l_4,0) \land \chk(\{l_0,l_3\},1) \land \chk(\{l_1,l_2,l_5,l_6\},2) \to \move(l_3)$
\State Rule63: $\bud=l_0\land \chk(\{l_3,l_4\},0) \land \chk(\{l_1,l_2,l_5\},1) \land \chk(l_6,2) \to \move(l_2)$
\State Rule64: $\bud=l_4\land \chk(\{l_2,l_5\},0) \land \chk(\{l_4,l_6\},1) \land \chk(\{l_0,l_1,l_3\},2) \to \move(l_4)$
\State Rule65: $\bud=l_0\land \chk(\{l_1,l_4,l_5\},0) \land \chk(l_6,1) \land \chk(\{l_2,l_3\},2) \to \move(l_1)$
\State Rule66: $\bud=l_0\land \chk(\{l_3,l_4,l_5\},0) \land \chk(\{l_2,l_6\},1) \land \chk(l_1,2) \to \move(l_2)$
\State Rule67: $\bud=l_1\land \chk(\{l_4,l_5\},0) \land \chk(\{l_0,l_1,l_3,l_6\},1) \land \chk(l_2,2) \to \move(l_1)$
\State Rule68: $\bud=l_2\land \chk(\{l_3,l_4\},0) \land \chk(\{l_0,l_2\},1) \land \chk(\{l_1,l_5,l_6\},2) \to \move(l_2)$
\State Rule69: $\bud=l_6\land \chk(\{l_1,l_2,l_4\},0) \land \chk(\{l_0,l_3,l_6\},1) \land \chk(l_5,2) \to \move(l_6)$
    \end{algorithmic}
\end{algorithm}


\subsection{Correctness of the Proposed Algorithm}
The proof of the correctness of the proposed algorithm
is very challenging because it includes many rules.
However, in this problem, 
the number of \PR~is fixed to 7,
all \PR s are initially in \SH~state,
and only \emph{connected} initial configurations are allowed;
this implies that 
there exist only the constant number of initial configurations.
Therefore, we implement a simulator (provided at \cite{git})
for the proposed algorithm 
which can generate all possible initial configurations;
there exist 3,652 initial configurations (as a result from the simulator).
Note that the proposed algorithm is deterministic 
and we assume an $\FSYNC$ scheduler,
thus there can be only one execution appears
when an initial configuration is given.
We had checked the proposed algorithm 
in every initial configuration using the simulation,
and the proposed algorithm solves the 7-pairbots-gathering problem
from all possible initial configurations.
Therefore, the following theorem holds.

\begin{theorem}
    The proposed algorithm solves the 7-pairbots-gathering problem
    from any arbitrary connected configuration 
    under an $\FSYNC$ scheduler.
\end{theorem}

\section{Conclusion}
\label{sec:conclude}
In this study, we have introduced a new computational model, the \PRM, 
consisting of paired robots that are called \PR s, 
which is based on the \emph{LCM} model \cite{SY99}.
In the proposed model, each \PR\ repeatedly changes the positional relation of 
the two robots (\ie \LN\ and \SH) to achieve the goal.
We presented 
the \emph{perpetual marching} and the \emph{7-pairbots-gathering} problems
to help to understand the computational power of the \PRM.
In particular, from the \emph{perpetual marching} problem,
we can clarify the difference of the computational power 
in terms of \emph{a scheduler};
the problem is solvable by 3 \PR s under an $\ASYNC$ scheduler,
but unsolvable by 6 or less LCM-robots even under an $\SSYNC$ scheduler.
Moreover, in the \emph{7-pairbots-gathering} problem,
we showed the difference of the computational power 
in terms of \emph{a visibility range};
the problem is solvable by \PR s with visibility range 1,
but unsolvable by LCM-robots with the same visibility range 
(visibility range 2 is necessary to solve).

The \PRM\ basically has similar variations of assumptions 
(\eg scheduler, geometric agreement, visibility, \etc), 
but it has one big different feature: 
every robot has one \emph{implicitly distinguishable} robot as its \bd. 
We introduced only two problems here, 
but we are considering various problems in the \PRM\ 
such as pattern formation (a line, a triangle, or a hexagon), 
filling problem, and uniform deployment, as the future work. 
Especially, we are interested in the problems 
which have been solved in other computational models
such as \emph{Amoebot}, \emph{SILBOT}, and \emph{MOBLOT}
to clarify the difference between \PRM~and these computational models for PM.


The \PR\ and the conventional \emph{LCM}-models have many common features, 
thus, we can consider the simulation of the \PRM\ 
using the \emph{LCM} model with some additional capabilities
(\eg light \cite{light1}).
Clarifying the minimum required capabilities for the \emph{LCM} model to simulate the \PRM\ is another future work.

\bibliographystyle{unsrtnat}
\bibliography{pairbot-arxiv-ver2}






\end{document}